\documentclass{article}
\usepackage{amsmath}
\usepackage{graphicx}
\usepackage{subfigure}
\usepackage{authblk}
\usepackage{hyperref}
\usepackage{etoolbox}
\usepackage{lmodern}
\usepackage{abstract}
\usepackage[T1]{fontenc}
\usepackage[utf8]{inputenc}
\usepackage{amsmath}
\usepackage{mathtools}
\usepackage{graphicx}
\usepackage{subfigure}
\usepackage{algorithm}
\usepackage{algorithmic}
\usepackage{tikz}
\usepackage{authblk}
\usepackage{hyperref}
\usepackage{etoolbox}
\usepackage{float}
\usepackage{lmodern}
\usepackage[T1]{fontenc}
\usepackage{braket}
\usepackage[utf8]{inputenc}
\usepackage[font=small,labelfont=bf]{caption} 
\usepackage{geometry}
\usepackage[font=small,labelfont=bf]{caption} 
\usepackage{geometry}
\newcommand\x[1]{\sigma_x^{#1}}
\newcommand\y[1]{\sigma_y^{#1}}
\newcommand\z[1]{\sigma_z^{#1}}\title{Electronic Structure Calculations and the Ising Hamiltonian}
\author[1]{Rongxin Xia}
\author[1]{Teng Bian}
\affil[1]{Department of Physics, Purdue University, West Lafayette, IN, 47907 USA}
\author[1,2,3,4]{Sabre Kais \thanks{kais@purdue.edu}}
\affil[2]{Department of Chemistry and Birck Nanotechnology Center, Purdue University,
West Lafayette, IN 47907 USA}
\affil[3]{Qatar Environment and Energy Research Institute, HBKU, Doha, Qatar}
\affil[4]{Santa Fe Institute,  1399 Hyde Park Rd, Santa Fe, NM 87501}
\date{}
\begin{document}
\maketitle
\vspace{-8ex}
\begin{abstract}
{\bf The exact solution of the Schr\"odinger equation for atoms, molecules and extended systems continues to be a "Holy Grail" problem for the field of atomic and molecular physics since inception. Recently, breakthroughs have been made in the development of hardware-efficient quantum optimizers and coherent Ising machines capable of simulating hundreds of interacting spins through an Ising-type Hamiltonian. One of the most vital questions associated with these new devices is: "Can these machines be used to perform electronic structure calculations?" In this study, we discuss the general standard procedure used by these devices and show that there is an exact mapping between the electronic structure Hamiltonian and the Ising Hamiltonian. The simulation results of the transformed Ising Hamiltonian for H$_2$, He$_2$, HeH$^+$, and LiH molecules match the exact numerical calculations. This demonstrates that one can map the molecular Hamiltonian to an Ising-type Hamiltonian which could easily be implemented on currently available quantum hardware. }
\end{abstract}

The determination of solutions to the Schr\"odinger equation is fundamentally difficult as the dimensionality of the corresponding Hilbert space increases exponentially with the number of particles in the system, requiring a commensurate increase in computational resources. Modern quantum chemistry --- faced with difficulties associated with solving the Schr\"odinger equation to chemical accuracy ($\sim$1 kcal/mole) --- has largely become an endeavor to find approximate methods.  A few products of this effort from the past few decades include methods such as: \emph{ab initio}, Density Functional, Density Matrix, Algebraic, Quantum Monte Carlo and Dimensional Scaling\cite{herschbach2012dimensional,iachello1995algebraic,kais_book,szabo1989modern}. However, all methods hitherto devised face the insurmountable challenge of escalating computational resource requirements as the calculation is extended either to higher accuracy or to larger systems. Computational complexity in electronic structure calculations\cite{PCCP,Frank,whitfield2014np} suggests that these restrictions are an inherent difficulty associated with simulating quantum systems.

Electronic structure algorithms developed for quantum computers provide a new promising route to advance the field of electronic structure calculations for large systems\cite{o2015scalable,Seth}. Recently, there has been an attempt at using an adiabatic quantum computing model --- as is implemented on the D-Wave machine --- to perform electronic structure calculations\cite{Alan-SR}. The fundamental concept behind the adiabatic quantum computing (AQC) method is to define a problem Hamiltonian, $H_P$, engineered to have its ground state encode the solution of a corresponding computational problem. The system is initialized in the ground state of a beginning Hamiltonian, $H_B$, which is easily solved classically.  The system is then allowed to evolve adiabatically as: $H(s)=(1-s) H_B+s H_P$ (where $s$ is a time parameter, $s\in[0,1]$). The adiabatic evolution is governed by the Schr\"odinger equation for the time-dependent Hamiltonian $H(s(t))$.

The largest scale implementation of AQC to date is by D-Wave Systems\cite{Boixo,Rose}. In the case of the D-Wave device, the physical process undertaken which acts as an adiabatic evolution is more broadly called \emph{quantum annealing} (QA). The quantum processors manufactured by D-Wave are essentially a transverse Ising model with tunable local fields and coupling coefficients.  The governing Hamiltonian is given as: $ H = \sum_i\Delta_i \sigma_x^i + \sum_{i}h_i \sigma_z^i + \sum_{i,j}J_{ij}\sigma_z^i \sigma_z^j$; where the parameters $\Delta_i$, $h_i$ and $J_{ij}$ are the physically tunable field, self-interaction and site-site interaction. The qubits are connected in a specified graph geometry, permitting the embedding of arbitrary graphs. Zoller and coworker presented a scalable architecture with full connectivity, which can be implemented with only local interactions\cite{Zoller}. The adiabatic evolution is initialized at $ H_B = -h\sum_i \sigma_x^i$ and evolves into the problem Hamiltonian: $\ H_P=\sum_{i}h_i \sigma_z^i + \sum_{i,j}J_{ij} \sigma_z^i \sigma_z^j$. This equation describes a classical Ising model whose ground state is --- in the worst case --- \textsc{NP}-complete. Therefore any combinatorial optimization \textsc{NP}-hard problem may be encoded into the parameter assignments, $\{h_i,J_{ij}\}$, of $H_P$ and may exploit the adiabatic evolution under $ H(s)=(1-s) H_B+s H_P$ as a method for reaching the ground state of $H_P$. More recently, an optically-based coherent Ising machine was developed; this machine is capable of finding the ground state of an Ising Hamiltonian populated by hundreds of coupled spin-1/2 particles\cite{Ising-1, Ising-2,Ising-3}. These challenging NP-hard problems are characterized by the difficulty in devising a polynomial-time algorithm, therefore solutions cannot be easily found using classical numerical algorithms in a reasonable time for large system sizes ($N$)\cite{Ising-1,Ising-2,Ising-3}. These special purpose machines may help in finding the solutions to some of the hardest problems in computing.

The technical scheme for performing electronic structure calculations on such an Ising-type machine can be summarized in the following four steps: First, write down the electronic structure Hamiltonian via the second quantization method in terms of creation and annihilation fermionic operators; Second, use the Jordan Wigner or the Bravyi-Kitaev transformation to move from fermionic operators to spin operators\cite{bravyi2002fermionic}; Third, reduce the Spin Hamiltonian which is a k-local in general to a 2-local Hamiltonian. Finally, map the 2-local Hamiltonian to an Ising-type Hamiltonian. 

Explicitly, this general procedure begins with a second quantization description of a fermionic system in which $N$ single-particle states can be either empty or occupied by a spineless fermionic particle\cite{mcweeny1969methods,szabo1989modern}. One may then use the tensor product of individual spin orbitals written as $|f_{0}...f_n \rangle$ to represent states in fermionic systems, where $f_j \in \left\{ 0,1 \right\}$ is the occupation number of orbital $j$. Any interaction within the fermionic system can be expressed in terms of products of the creation and annihilation operators $a_j^{\dagger}$ and $a_j$, for $j \in \left\{0, ..., N\right\}$. Thus, the molecular electronic Hamiltonian can be written as:

\begin{equation}
\hat H= \sum_{i,j}h_{ij}a_i^\dagger a_j+\frac{1}{2}\sum_{i,j,k,l} h_{ijkl}a_i^\dagger a_j^\dagger a_ka_l.
\end{equation}

The above coefficients $h_{ij}$ and $h_{ijkl}$ are one and two-electron integrals which can be precomputed through classical methods and are used as inputs for the quantum simulation. The next step is to convert to a Pauli matrix representation of the creation and annihilation operators. We can then use the Bravyi-Kitaev transformation or the Jordan-Wigner transformation\cite{bravyi2002fermionic, seeley2012bravyi} to map between the second quantization operators and Pauli matrices $\left\{\sigma_x,\sigma_y,\sigma_z\right\}$. The molecular Hamiltonian takes the general form:

\begin{equation}
H=\sum\limits_{i,\alpha}h_\alpha^i\sigma_\alpha^i+\sum\limits_{ij\alpha\beta}h_{\alpha\beta}^{ij}\sigma_\alpha^i\sigma_\beta^j+\sum\limits_{ijk\alpha\beta\gamma}h_{\alpha\beta\gamma}^{ijk}\sigma_\alpha^i\sigma_\beta^j\sigma_\gamma^k+...
\end{equation}
Within the above, the indices $\alpha={x,y,z}$ are anisotropic directions and the indices $i$ and $j$ are for the spin orbitals. Now, after having developed a $k$-local spin Hamiltonian (many-body interactions), one should use a general procedure \cite{Ref-7,Ref-14} to reduce to a 2-local (two-body interactions) spin Hamiltonian; this is a requirement since the proposed experimental systems are typically limited to restricted forms of two-body interactions. Therefore, universal adiabatic quantum computation requires a method for approximating a quantum many-body Hamiltonian up to an arbitrary spectral error using at most two-body interactions. Hamiltonian gadgets, for example offer a systematic procedure to address this requirement. Recently, we have employed analytical techniques resulting in a reduction of the resource scaling as a function of spectral error for the most commonly used device classifications: three- to two-body and k-body gadgets\cite{Kais-2015}.

Here we present a universal way of mapping an $n$ qubit Hamiltonian, $H$, which depends on $\sigma_x$, $\sigma_y$, $\sigma_z$ to an $rn$ qubits Hamiltonian, $H'$, consisting of only product of $\sigma_z$. In this process we increase the number of qubits from $n$ to $r n$, where the integer $r$ plays the role of a "variational parameter" to achieve the desired accuracy in the final step of energy calculations.

If the eigenstate of $H$ is given by $\psi=\sum_i a_i \phi_i$, then the spin operators
($\sigma_x^i$, $\sigma_y^i$, $\sigma_z^i$) 
acting on the $i^{th}$ qubit in $|\psi\rangle$. $H$
can be transformed to $H'$ in the newly mapped space, to state space $|\Psi\rangle$ from $|\psi\rangle$. The mapping state $|\Psi\rangle$ includes the $r$ copies of the original 
$n$ qubits. The mapping of Hamiltonian from $H$ to $H'$ can be written as: 

\begin{equation}
\begin{aligned}
&\x{i} \rightarrow \frac{1-\z{i_j}\z{i_k}}{2}S'(j)S'(k) \qquad \y{i} \rightarrow {\bf i}\frac{\z{i_k}-\z{i_j}}{2}S'(j)S'(k)\qquad\\ &\z{i} \rightarrow \frac{\z{i_j}+\z{i_k}}{2}S'(j)S'(k)\qquad I^i \rightarrow \frac{1+\z{i_j}\z{i_k}}{2}S'(j)S'(k) 
\end{aligned}
\end{equation}

$\sigma^{i_j}$ implies mapping the $i^{th}$ qubit in $|\psi\rangle$ to the $j^{th}$ qubit of the $n$-qubit space $|\Psi\rangle$ (or the spin operator $\sigma_z$ acting on the $[(j-1)n+i]$ qubit) and $S'(j), S'(k)$ represent the sign of the $j^{th}, k^{th}$ of the $n$ qubits in the new state of the $r n$ qubits. Appropriately accounting for the correct signs guarantees that the number of $n$ qubits in the new basis may be only positive. If we count the total number of $n$ qubits in the basis $\phi_i$ as $b_i$, we can relate this number to the coefficient $a_i$ of basis $\phi_i$ in original normalized state of $n$ qubits by $\frac{S(b_i)b_i}{\sqrt{\sum_{m}b_m^2}}=a_i$ where $S(b_i)$ represents the sign associated with $b_i$. Now, $H'$ includes only products of $\sigma_z$ and $I$, which is diagonal.  Details of the transformation are presented as examples in the Supplementary Materials

\begin{center}
\includegraphics[height=2in,]{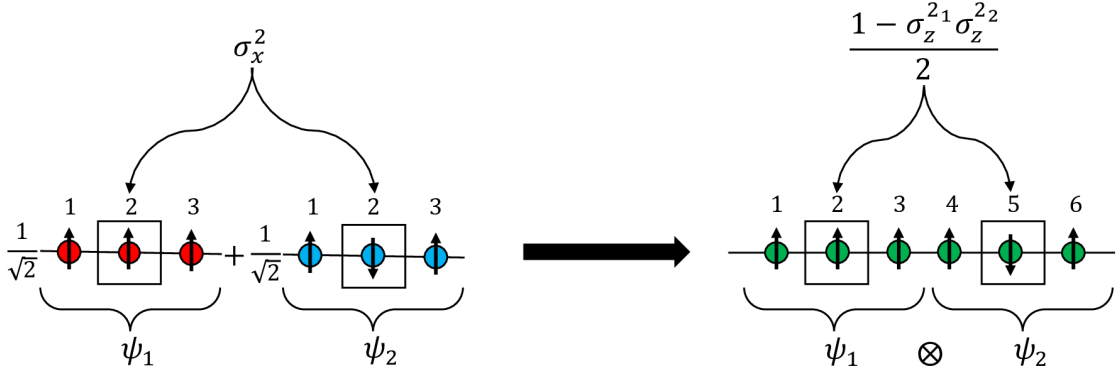}
\captionof{figure}{A schematic representation of the mapping of $\x{2}$ between different basis from a state $|\psi\rangle=\frac{1}{\sqrt{2}}|000\rangle+\frac{1}{\sqrt{2}}|010\rangle$ to state $|\Psi\rangle=|000010\rangle$ where $0$ represents spin up and $1$ represents spin down. The spin operators act on the second qubit in the original Hamiltonian basis and on the second and fifth qubits 
in the mapped Hamiltonian basis.} 
\end{center}

So far we are able to transform our Hamiltonian to a $k$-local Hamiltonian including only products of $\sigma_z$ terms. In the Supplementary Materials we show the details of the transformation to 2-local spin Hamiltonian but here we present an example of transforming 3-local to a 2-local Ising Hamiltonian\cite{bian2013experimental}.

\begin{equation}
min(\pm x_1x_2x_3)=min(\pm x_4x_3+x_1x_2-2x_1x_4-2x_2x_4+3x_4)\ \ x_1,x_2,x_3,x_4 \in \{0,1\}.
\end{equation}
Here, we see that by including $x_4$, one can show that minimizing 3-local is equivalent to minimizing the sum of 
2-local terms. 

Finally, we succeed in transforming our initial complex electronic structure Hamiltonian from the second-quantization form to an Ising-type Hamiltonian which can be solved using 
existing quantum computing hardware \cite{Ising-1,Jay,o2015scalable,Chris}. 

To illustrate this proposed method (details are in the Supplementary Materials), we present calculations for the Hydrogen molecule H$_2$, the Helium dimer He$_2$, HeH$^+$ diatomic molecule and the LiH molecule. First, we used the Bravyi-Kitaev transformation and the Jordan-Wigner transformation to convert the diatomic molecular Hamiltonian
in the minimal basis set (STO-6G) to the spin Hamiltonian of ($\sigma_x$, $\sigma_y$, $\sigma_z$). Then we used our transformed Hamiltonian in $r n $-qubit space to obtain a diagonal $k$-local Hamiltonian of $\sigma_z$ terms. Finally, we reduced the locality 
to get a 2-local Ising Hamiltonian of the general form:

\begin{equation}
H'=\sum_i h'_{i}\z{i}+\sum_{ij} J'_{ij}\z{i}\z{j} .
\end{equation}

\begin{figure}[h]
\begin{minipage}[t]{0.5\linewidth}
\centering
\includegraphics[width=3.5in]{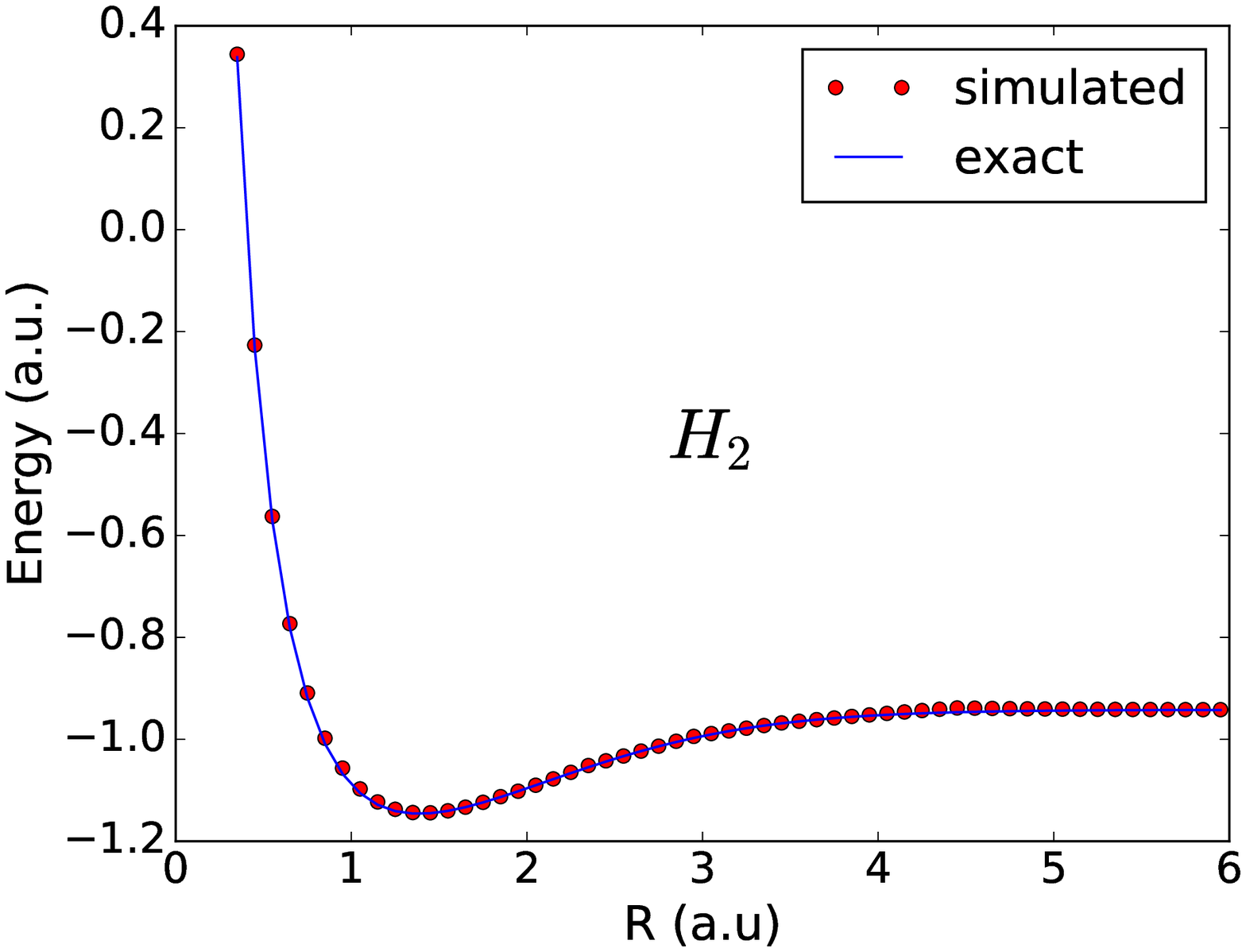}
\caption*{(a)}
\label{fig:side:a}
\end{minipage}%
\begin{minipage}[t]{0.5\linewidth}
\centering
\includegraphics[width=3.5in]{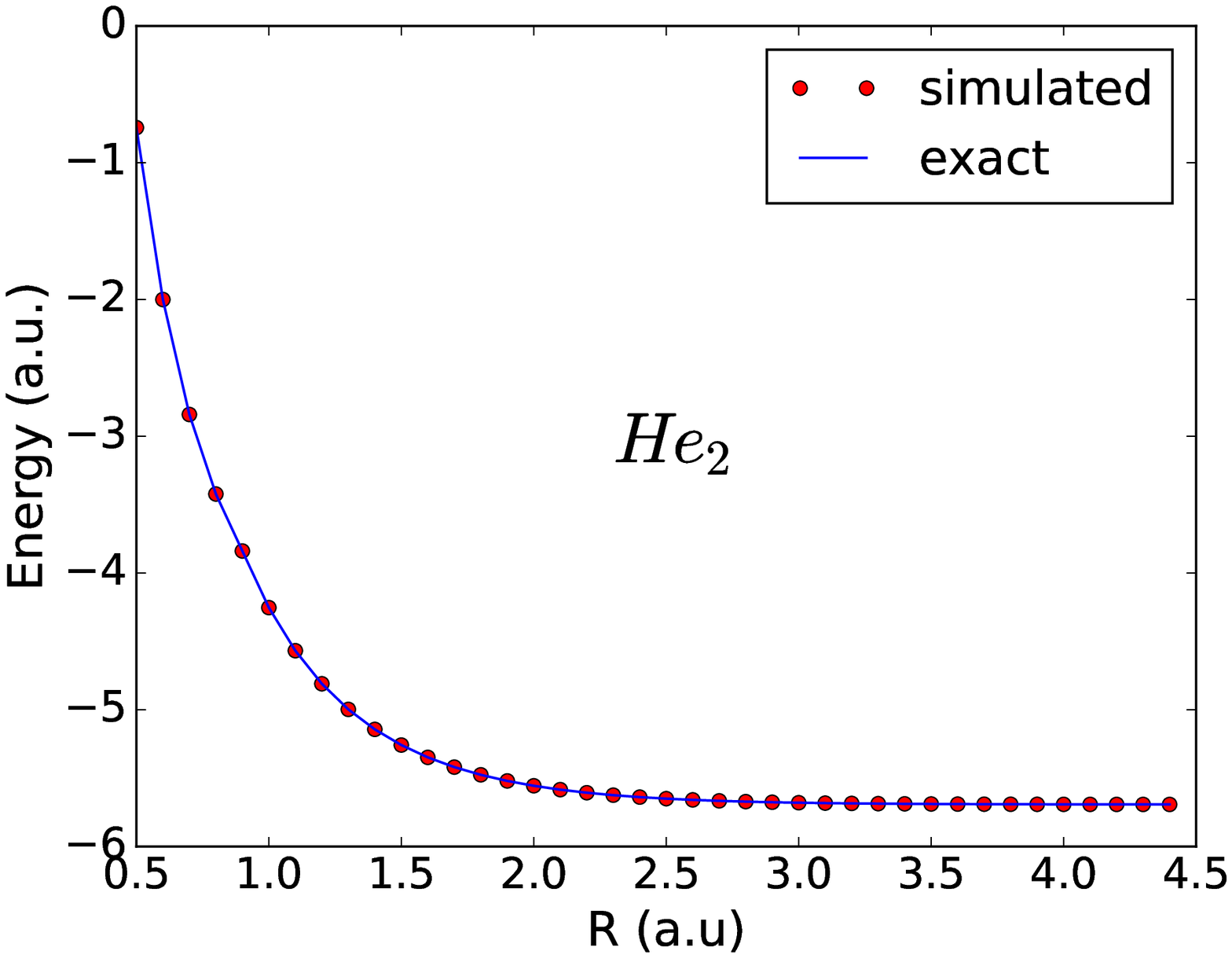}
\caption*{(b)}
\label{fig:side:b}
\end{minipage}
\begin{minipage}[t]{0.5\linewidth}
\centering
\includegraphics[width=3.5in]{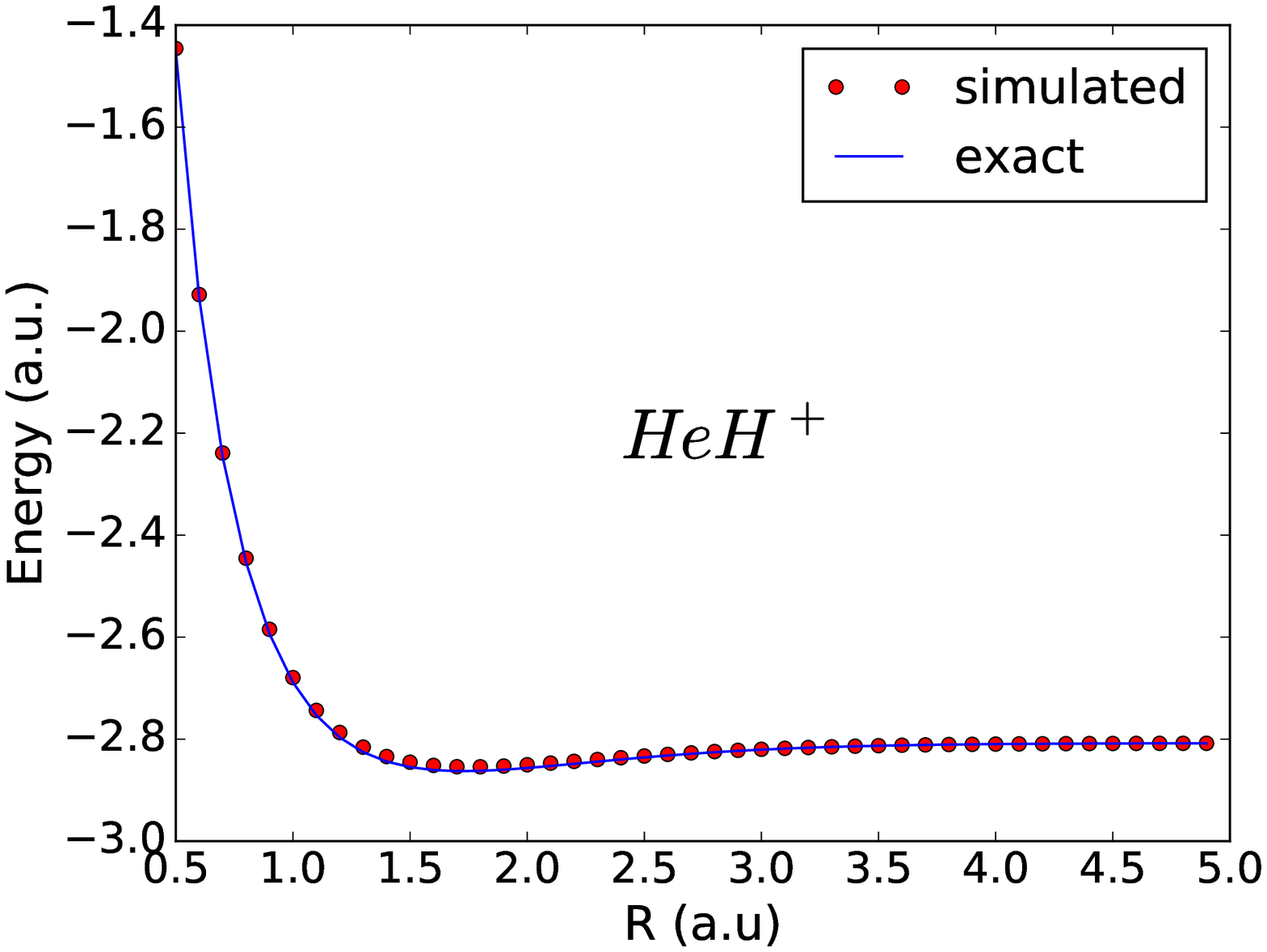}
\caption*{(c)}
\label{fig:side:c}
\end{minipage}
\begin{minipage}[t]{0.5\linewidth}
\centering
\includegraphics[width=3.5in]{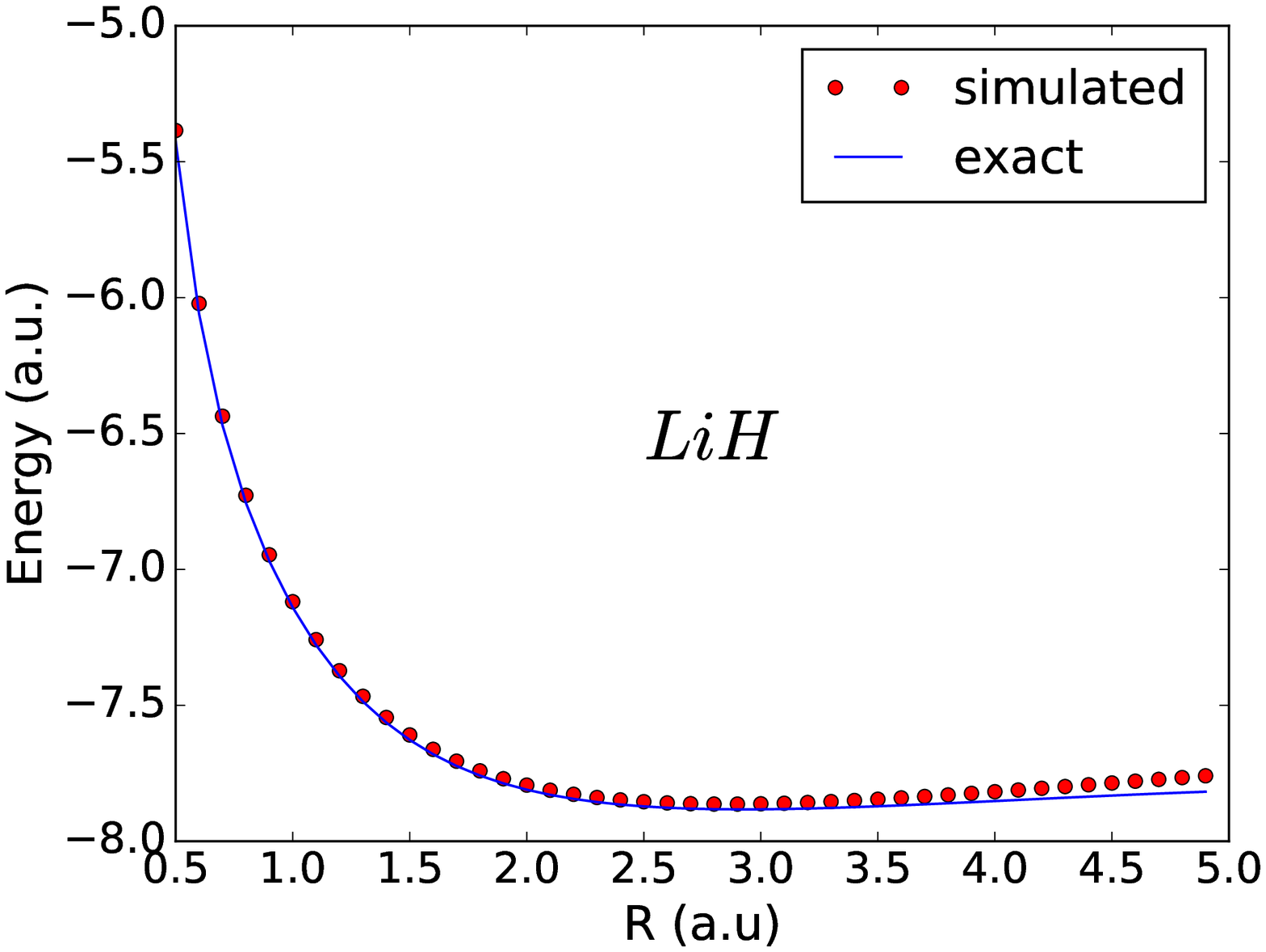}
\caption*{(d)}
\label{fig:side:d}
\end{minipage}
\caption{A comparison of numerical results of ground state energy of the Ising Hamiltonian with the exact STO-6G calculations of the ground state of H$_2$, He$_2$, HeH$^+$ and LiH molecules as one varies the internuclear distance $R$.}
\end{figure}

These results show that our simulations based on a transformed Ising-tpe Hamiltonian matche the exact result for these diatomic molecules. This demonstrates that one can generally map the electronic ground state energy of a molecular Hamiltonian to an Ising-type Hamiltonian which could easily be implemented on presently available quantum hardware. Moreover, the recent experimental results for simple few electrons diatomic molecules presented by the IBM group have shown that a hardware-efficient optimizer implemented on a 6-qubit superconducting quantum processor is capable of producing the potential energy surfaces of such molecules\cite{Jay}.
The development of efficient quantum hardware and the possibility of mapping the electronic structure problem into an Ising-type Hamiltonian may grant efficient ways to obtain exact solutions to the Schr\"odinger equation, this being one of the most daunting computational problems present in both chemistry and physics.

\section*{Acknowledgement}

We would like to thank Dr. Ross  Hoehn for critical reading of the manuscript.

\bibliographystyle{unsrt}
\bibliography{cite12.bib}

\newpage
\begin{center}
\LARGE{Supplementary Material}
\end{center}

\subsection*{Mapping between Hamiltonian}
Here, we present a procedure to construct a diagonal Hamiltonian with a minimum eigenvalue corresponding to the ground state of a given initial Hermitian Hamiltonian.

For a given initial Hermitian Hamiltonian, an eigenstate $|\psi\rangle$ can be expanded in a basis set $|\phi_i\rangle$ as $|\psi\rangle=\sum_i a_i|\phi_i\rangle$. This basis set consists of different combinations of spin-up and -down qubits.  First we will assume that all expansion coefficients, $a_i$, are nonnegative and will map the state to a new state $|\Psi\rangle$ according to the following rules:

\begin{itemize}
\item $|\Psi\rangle$ can be written as $|\Psi\rangle=\otimes_i\otimes_{j=1}^{b_i}|\phi_i\rangle$ and $r=\sum_ib_i$.
\item If the original state $|\psi\rangle$ exists within an $n$-qubit subspace, the new state $|\Psi\rangle$ should be in an $r n$ qubit space, where $r$ is the number of times we must replicate the $n$ qubits to a achieve an arbitrary designated accuracy.
\item The number of times, $b_i$, we repeat the basis $|\psi\rangle$ in an $rn$ qubit state, $|\Psi\rangle$, approximates $a_i$ by $\frac{b_i}{\sqrt{\sum_mb_m^2}}$. If $r$ is large enough, $b_i$ is proportional to $a_i$ or we can just view $a_i \approx \frac{b_i}{\sqrt{\sum_m b_m^2}}$, where $\sqrt{\sum_m b_m^2}$ is the normalization factor.
\end{itemize}

Here we introduce notation to be used throughout the remaining text:

{\bf Notation 1:} We designate the $i^{th}$ qubit within the $k^{th}$ $n$ qubit subspace of the $rn$ state space of $|\Psi\rangle$ as $i_k$. 

{\bf Notation 2:}: We use $b(j)$ to represent the $j^{th}$ $n$-qubit in the space of $|\Psi\rangle$, is in the basis $|\phi_{b(j)}\rangle$.

$b_i$ may only be non-negative, yet $a_i$ may be positive or negative. For a negative 
$a_i$, we will introduce a function $S(b_i)$  containing the sign information to account for $b_i$ being non-negative.  The mapping is described by the following rules:
\begin{itemize}
\item $S(b_i)$ is the sign associated with $b_i$ coefficient which is negative if $a_i$ is negative and is positive if $a_i$ is positive.
\item For the $rn$ qubit state $|\Psi\rangle$, we can use a function $S'(i)$ to record the sign associated with each $n$ qubits in the $rn$ qubits space. $S'(i)$ represent the sign of the $i^{th}$ $n$ qubits in the $rn$ qubits space. Thus, $b_i=|\sum_{|\phi_{b(j)}\rangle=|\phi_i\rangle}S'(j)|$ and $S(b_i)$ is the sign of $\sum_{|\phi_{b(j)}\rangle=|\phi_i\rangle}S'(j)$.
\item As before, $b_i$ is the integer which approximates $a_i$ by $\frac{b_iS(b_i)}{\sqrt{\sum_mb_m^2}}$. If $r$ is large enough, we can just view $a_i \approx \frac{b_iS(b_i)}{\sqrt{\sum_m b_m^2}}$. 
\item $|\Psi\rangle$ can be written as $|\Psi\rangle=\otimes_i\otimes_{j=1}^{b_i}|\phi_i\rangle$ and $r=\sum_{i=1}^{r}|S'(i)|$.
\end{itemize}

\begin{figure}
\begin{center}
\includegraphics[height=1in,]{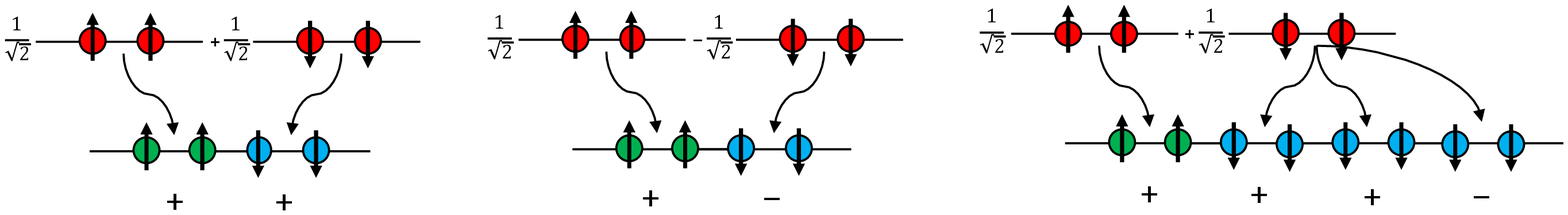}
\caption{
{\bf Left}: If the $2$ qubits state is $|\psi\rangle =\frac{1}{\sqrt{2}}|00\rangle+\frac{1}{\sqrt{2}}|11\rangle$ then, the $4$ qubits state is $|\Psi\rangle =|0011\rangle$ with $S'(1)=S'(2)=1$, $r=2$, $b_1=b_2=1$ and $S(b_1)=S(b_2)=1$.
{\bf Middle:} If the $2$ qubits state is $|\psi\rangle =\frac{1}{\sqrt{2}}|00\rangle-\frac{1}{\sqrt{2}}|11\rangle$ then, the $4$ qubits state is $|\Psi\rangle =|0011\rangle$ with $S'(1)=-S'(2)=1$, $r=2$, $b_1=b_2=1$ and $S(b_1)=-S(b_2)=1$.
{\bf Right:} If the $2$ qubits state is $|\psi\rangle =\frac{1}{\sqrt{2}}|00\rangle-\frac{1}{\sqrt{2}}|11\rangle$ then, the $8$ qubits state is $|\Psi\rangle= |00111111\rangle$ with $S'(1)=S'(2)=S'(3)=-S'(4)=1$, $r=4$, $b_1=b_2=1$ and $S(b_1)=S(b_2)=1$. ($S'(3)$ and $S'(4)$ cancel out and thus $S(b_2)=1$).}
\end{center}
\end{figure}

{\bf Theorem 1:} With the mapping between $|\psi\rangle$ and $|\Psi\rangle$ as described in above, we can find a mapping between the Hamiltonian in the space of $|\psi\rangle$ to the space of $|\Psi\rangle$.

{\bf Theorem 2:} $\langle\phi_{b(j)}|\otimes_i I_i|\phi_{b(k)}\rangle$ is equal to $\langle\Psi|\prod_i\frac{1+\z{i_j}\z{i_k}}{2}|\Psi\rangle$, which means $I_i$ in the space of $|\psi\rangle$ can be mapped to $\frac{1+\z{i_j}\z{i_k}}{2}$ in the space of $|\Psi\rangle$.

{\bf Proof:} Clearly, $I_i$ in the space of $|\psi\rangle$ is to observe if $i^{th}$ digits of $|\phi_{b(j)}\rangle$ and $|\phi_{b(k)}\rangle$ are the same or not. If they are the same, it yields 1, otherwise 0.  On the other hand, $\frac{1+\z{i_j}\z{i_k}}{2}$ in the space of $|\Psi\rangle$ is to check the $i^{th}$ digits of the  $j^{th}$ $n$-qubits subspace ($|\phi_{b(j)}\rangle$) and the $k^{th}$ $n$-qubit subspace ($|\phi_{b(k)}\rangle$) are the same or not. If they are the same it yields 1, otherwise 0. (For $\frac{1-\z{i_j}\z{i_k}}{2}$ we omit the operators for other digits which are the identity $I$.)

Thus we get:

\begin{equation}
\langle\phi_{b(j)}|\otimes I_i|\phi_{b(k)}\rangle =\langle\Psi|\prod_i\frac{1+\z{i_j}\z{i_k}}{2}|\Psi\rangle .
\end{equation}

If and only if all digits of $|\phi_{b(j)}\rangle$ and $|\phi_{b(k)}\rangle$ are the same, the left and right results are equal to 1 otherwise they are equal to 0. 

{\bf Theorem 3:}$\langle\phi_{b(j)}|\otimes_{i<m} I_i\otimes \x{m}\otimes_{i>m}I_i|\phi_{b(k)}\rangle$ is equal to $\langle\Psi|\prod_{i<m}\frac{1+\z{i_j}\z{i_k}}{2}\times\frac{1+\z{m_j}\z{m_k}}{2}\times\prod_{i>m}\frac{1-\z{i_j}\z{i_k}}{2}|\Psi\rangle$, which means $\x{i}$ in the space of $|\psi\rangle$ can be mapped as $\frac{1-\z{i_j}\z{i_k}}{2}$ in the space of $|\Psi\rangle$.

{\bf Proof:} Clearly, $\x{i}$ in the space of $|\psi\rangle$ is to check if the $i^{th}$ digit of $|\phi_{b(j)}\rangle$ and $|\phi_{b(k)}\rangle$ are the same or not. If they are the same it yields 0, otherwise 1.  Similarly, $\frac{1-\z{i_j}\z{i_k}}{2}$ in the space $|\Psi\rangle$ is to verify the $i^{th}$ digits of  the $j^{th}$ $n$-qubit subspace ($|\phi_{b(j)}\rangle$) and the $k^{th}$ $n$-qubit subspace ($|\phi_{b(k)}\rangle$) are identical. If they are the same it gives 0, otherwise 1. (For $\frac{1-\z{i_j}\z{i_k}}{2}$ we omit operators for other digits which are the identity $I$.)

Thus we get:

\begin{equation}
\langle\phi_{b(j)}|\otimes_{i<m} I_i\otimes \x{m}\otimes_{i>m}I_i|\phi_{b(k)}\rangle= \langle\Psi|\prod_{i<m}\frac{1+\z{i_j}\z{i_k}}{2}\times\frac{1+\z{m_j}\z{m_k}}{2}\times\prod_{i>m}\frac{1-\z{i_j}\z{i_k}}{2}|\Psi\rangle .
\end{equation}

 $\x{i}$ and $\frac{1-\z{i_j}\z{i_k}}{2}$ have the  same function in different space to check the digits of $|\phi_{b(j)}\rangle$ and $|\phi_{b(j)}\rangle$.

Also, $\y{i}$ and ${\bf i}\frac{\z{i_k}-\z{i_j}}{2}$ have the same function in different spaces.  These operators are used to check the $i^{th}$ digits of $|\phi_{b(j)}\rangle$ and $|\phi_{b(j)}\rangle$. Also, $\z{i}$ and $\frac{\z{i_j}+\z{i_k}}{2}$ have the  same function in different spaces to check the $i^{th}$ digits of $|\phi_{b(j)}\rangle$ and $|\phi_{b(j)}\rangle$. This can be easily verified by the above discussion.

{\bf Theorem 4:} Any Hermitian Hamiltonian in the space of $|\psi\rangle$ can be written in the form of Pauli and Identity Matrices, which can be mapped to the space of $|\Psi\rangle$ as  described above.

{\bf Notation 3:} We denote the mapping between the $j_{th}$ $n$-qubit subspace and the $k^{th}$ $n$-qubit subspace in $|\Psi\rangle$, $\frac{1-\z{i_j}\z{i_k}}{2}$ as $X_i^{(j,k)}$, $\frac{\z{i_j}+\z{i_k}}{2}$ as $Z_i^{(j,k)}$, ${\bf i}\frac{\z{i_k}-\z{i_j}}{2}$ as $Y_i^{(j,k)}$ and $\frac{1+\z{i_j}\z{i_k}}{2}$ as $I_i^{(j,k)}$.

{\bf Proof:}
If $H$ can be written as:

\begin{equation}
H=\otimes_a\x{a}\otimes_b\y{b}\otimes_c\z{c}\otimes_dI_{d} .
\end{equation}

We can write the mapped $H'_{(j,k)}$ as:

\begin{equation}
H'_{(j,k)}=\prod_aX_a^{(j,k)}\prod_bY_b^{(j,k)}\prod_cZ_c^{(j,k)}\prod_dI_d^{(j,k)} .
\end{equation}

It can be verified following the rules above that:

\begin{equation}
\langle\Psi|H'_{(j,k)}|\Psi\rangle=\langle\phi_{b(j)}|H|\phi_{b(k)}\rangle .
\end{equation}

Thus, if we add the sign functions $S'(j)$ and $S'(k)$, we achieve:

\begin{equation}
\begin{aligned}
&\langle\Psi|\sum_{j,k}^{j\neq k, j,k\leq r}H_{(j,k))}S'(j)S'(k)|\Psi\rangle=\sum_{j,k}^{j\neq k, j,k \leq r}\langle\phi_{b(j)}|H|\phi_{b(k)}\rangle S'(j)S'(k)\\
&=\sum_{j,k}^{j\neq k, j,k\leq 2^n} b_jS(b_j)b_kS(b_k)\langle\phi_j|H|\phi_k\rangle=\sum_mb_m^2\sum_{j,k}^{j\neq k, j,k\leq 2^n} a_ja_k\langle\phi_j|H|\phi_k\rangle .
\end{aligned}
\end{equation}

Also, in the same basis, we have:

\begin{equation}
\begin{aligned}
X_i^{(j,j)}=\frac{1-\z{i_j}\z{i_j}}{2}=0 \quad
Y_i^{(j,j)}={\bf i}\frac{\z{i_j}-\z{i_j}}{2}=0 \quad
Z_i^{(j,j)}=\frac{\z{i_j}+\z{i_j}}{2}=\z{i_j} \quad
I_i^{(j,j)}=\frac{1+\z{i_j}\z{i_j}}{2}=I
\end{aligned}
\end{equation}

Thus, as before, we can also get:
\begin{equation}
\langle\Psi|\sum_{j}^{j\leq r}H_{(j,j)}S'(j)S'(j)|\Psi\rangle=\sum_{j}^{j\leq r}\langle\phi_{b(j)}|H|\phi_{b(j)}\rangle S'(j)S'(j)=\sum_{j}^{j\leq 2^n} b_j^2\langle\phi_j|H|\phi_j\rangle=\sum_mb_m^2\sum_{j}^{j\leq
 2^n} a_j^2\langle\phi_j|H|\phi_j\rangle
\end{equation}

Combining the two together we can get:

\begin{equation}
\langle\Psi|\sum_{j,k}^{j,k\leq r}H_{(j,k)}S'(j)S'(k)|\Psi\rangle=\sum_mb_m^2\sum_{j,k}^{j,k\leq 2^n} a_ja_k\langle\phi_j|H|\phi_k\rangle
\end{equation}

We construct a matrix, $C$, in the space of $|\Psi\rangle$, which has elements $\sum_mb_m^2$ as:

\begin{itemize}
\item $C=\sum_\pm(\sum_i(\prod_{k=1_i}^{n_i}\frac{1\pm\z{k}}{2})S'(i))^2$
\item $\sum_\pm$ over all combination of positive and negative signs of each digit in each $i^{th}$ of the $n$-qubit in space $|\Psi\rangle$.
\item $\sum_i$ over all $n$-qubit collection in $|\Psi\rangle$ to check whether each $n$ qubits is in a certain state.
\item $\prod_{k=1_i}^{n_i}$  over each qubits of $i^{th}$ $n$ qubits in $|\Psi\rangle$ .
\item $\frac{1\pm\z{k}}{2}$ is to check whether $k^{th}$ qubits of $i^{th}$ $n$-qubit subspace is in a certain state. $\frac{1+\z{k}}{2}$ is 1 when the $k^{th}$ qubits is present in the basis $|0\rangle$, 0 otherwise. 
$\pm$ is to go over combination by $\sum_\pm$.
\end{itemize}

So far, we have established a mapping between $|\psi\rangle$ and $|\Psi\rangle$. The Hamiltonian $H$ in the space of $|\psi\rangle$ and $\sum_{(j,k)}^{j,k\leq r}H'_{(j,k)}$ in the space of $|\Psi\rangle$. Also we have constructed a matrix $C$ to compute $\sum_mb_m^2$ corresponding to  $|\Psi\rangle$. Thus, we have the final results:

\begin{equation}
\langle\Psi|\sum_{j,k}^{j,k\leq r}H_{(j,k)}S'(j)S'(k)|\Psi\rangle=\sum_mb_m^2\sum_{j,k}^{j,k\leq 2^n} a_ja_k\langle\phi_j|H|\phi_k\rangle
\end{equation}

Here, we present an algorithm combing $\sum_{(j,k)}^{j,k\leq r}H'_{(j,k)}$ and $C$ to calculate the ground state  of the initial Hamiltonian $H$.

{\bf Notation 4:} We mark the eigenvalue of $|\psi\rangle$ for $H$ as $\lambda'$. According to the relationship above, the eigenvalue of $|\Psi\rangle$ for $H'$ is $\sum_m b_m^2\lambda'$. Thus, if we choose a $\lambda$ and construct a Hamiltonian $H'-\lambda C$. The eigenvalue of $|\Psi\rangle$ for $H'-\lambda C$ is $\sum_mb_m^2(\lambda'-\lambda)$

\begin{algorithm}
\caption{}
\begin{algorithmic}
\FOR{$i$ from $0$ to $\lfloor\frac{r}{2}\rfloor$}
\STATE Set the signs of the first $i$ $n$ qubits to be negative and the others to be positive. Set $\lambda$ to be a large number (at least larger than the ground state value and this will avoid the case $\sum_mb_m^2=0$ ).
\STATE Construct $H'$ and $C$.
\WHILE{the ground eigenvalue of $(H'-\lambda C) < 0$}
\STATE Calculate $H'-\lambda C$ and get the state $|\Psi'\rangle$ and the corresponding eigenvalue $\sum_{m}b_m=^2(\lambda'-\lambda) $.
\STATE Calculate $C$ on $|\Psi'\rangle$ to get $\sum_mb_m^2$ then get $\lambda'$. Set $\lambda$ to be $\lambda'$.
\ENDWHILE
\STATE $\lambda'$ is the smallest eigenvalue with certain signs.
\ENDFOR
\STATE Compare all $\lambda'$ and the smallest one is the ground state energy of $H$.
\end{algorithmic}
\end{algorithm}

{\bf Theorem 5:} The algorithm above converges to the minimum eigenvalue of H by finite iterations.

{\bf Proof:} 

{\bf Monotonic Decreasing} If we can find an eigenstate $|\Psi\rangle$ of $H'-\lambda C$ with eigenvalue $\sum_m b_m^2 (\lambda'-\lambda) < 0$. Because $\sum_m b_m^2\geq 0$ we get $\lambda'-\lambda<0 $. This means we can find an eigenstate $|\psi\rangle$ of $H$ with an eigenvalue $\lambda'$ and $\lambda'-\lambda<0$. Thus each time $\lambda$ decrease monotonically.

{\bf The minimum eigenvalue:} Here we prove that the minimum eigenvalue of $H$ is achievable and the loop will converge when we obtain the minimum eigenvalue. According to {\bf Monotonic Decreasing}, each time the eigenvalue we get will decrease. Because we have finite number of eigenvalues, which means we will finally come to the minimum eigenvalue. Also, if we set $\lambda$ to be the minimum eigenvalue of $H$, $H'-\lambda C = \sum_m b_m^2 (\lambda'-\lambda)\geq 0$ because $\sum_m b_m^2\geq 0$ and $\lambda'-\lambda \geq 0$ Also, if $\lambda'$ is just the minimum eigenvalue, we get $\lambda'-\lambda=0$ and the loop stops.

Thus, we prove that the eigenvalue decreases and finally converges to the minimum eigenvalue of H.

{\bf Theorem 6:} To account for the sign, we just need to set $i$ from $0$ to $\lfloor\frac{r}{2}\rfloor$ and set signs of the first $i^{th}$ $n$-qubit to be negative and the others to be positive in $|\Psi\rangle$.

{\bf Proof:} If we have $n$ qubits in the $|\Psi\rangle$ space with negative sign, where $|\Psi\rangle$ has total $i$ $n$ qubits with negative sign. If this $n$ qubits are not in first $i$ $n$ qubits, we can rearrange it to the first $i$ $n$ qubits by exchanging it with $n$ qubits in first $i$ $n$ qubits which has positive sign. Thus all combination can be reduced to the combination stated in {\bf Theorem 6}.

Thus, we have established a transformation from an initial Hermitian Hamiltonian to a diagonal Hamiltonian and presented an algorithm to calculate the minimum eigenvalue of initial Hamiltonian using the diagonal Hamiltonian.

\subsection*{Example:}

To illustrate the above procedure, we give details of the transformation for the simple model of two spin-$\frac{1}{2}$ electrons with an exchange coupling constant $J$ in an effective transverse magnetic field of strength $B$. This simple model has been used to discuss the entanglement for H$_2$ molecule\cite{huang2005entanglement}. The general Hamiltonian for such a system is given by:

\begin{equation}
H=-\frac{J}{2}(1+\gamma)\x{1}\x{2}-\frac{J}{2}(1-\gamma)\y{1}\y{2}-B\z{1}-B\z{2} ,
\end{equation}

where $\gamma$ is the degree of anisotropy.

In the $\{|00\rangle,|10\rangle,|01\rangle,|11\rangle\}$ basis, the eigenvectors can be written as (here we just use the eigenvectors to show how we map the Hamiltonian but in actual calculation we do not know the eigenvectors):

\begin{equation}
\begin{aligned}
&|\xi_1\rangle=\frac{1}{\sqrt{2}}(|10\rangle+|01\rangle)\\
&|\xi_2\rangle=\frac{1}{\sqrt{2}}(|10\rangle-|01\rangle)\\
&|\xi_3\rangle=\sqrt{\frac{\alpha-2B}{2\alpha}}|11\rangle+\sqrt{\frac{\alpha+2B}{2\alpha}}|00\rangle\\
&|\xi_4\rangle=\sqrt{\frac{\alpha+2B}{2\alpha}}|11\rangle-\sqrt{\frac{\alpha-2B}{2\alpha}}|00\rangle ,
\end{aligned}
\end{equation}

where $\alpha=\sqrt{4B^2+J^2\gamma}$.

If we set $r=2$, for example, if $|\psi\rangle=|\xi_1\rangle$, $|\Psi\rangle=|10\rangle\otimes|01\rangle$ with $S'(1)=1$ and $S'(2)=1$. If $|\psi\rangle=|\xi_2\rangle$,
$|\Psi\rangle=|10\rangle\otimes|01\rangle$ with $S'(1)=-1$ and $S'(2)=1$.

Abiding the previous mapping, the mapped Hamiltonian $H'$ and matrix $C$ can be written as:

\begin{equation}
\begin{aligned}
&H'_{(1,1)}=-B\z{1}-B\z{2}\\
&H'_{(2,2)}=-B\z{3}-B\z{4}\\
&H'_{(1,2)}=-\frac{J}{2}(1+\gamma)\frac{1-\z{1}\z{3}}{2}\frac{1-\z{2}\z{4}}{2}-\frac{J}{2}(1-\gamma)\frac{\z{1}-\z{3}}{2}\frac{\z{4}-\z{2}}{2}\\
&-B\frac{\z{1}+\z{3}}{2}\frac{1+\z{2}\z{4}}{2}-B\frac{\z{2}+\z{4}}{2}\frac{1+\z{1}\z{3}}{2}\\
&H'_{(2,1)}= H'_{(1,2)} .\\
\end{aligned}
\end{equation}

\begin{figure}
\begin{center}
\includegraphics[height=2in,]{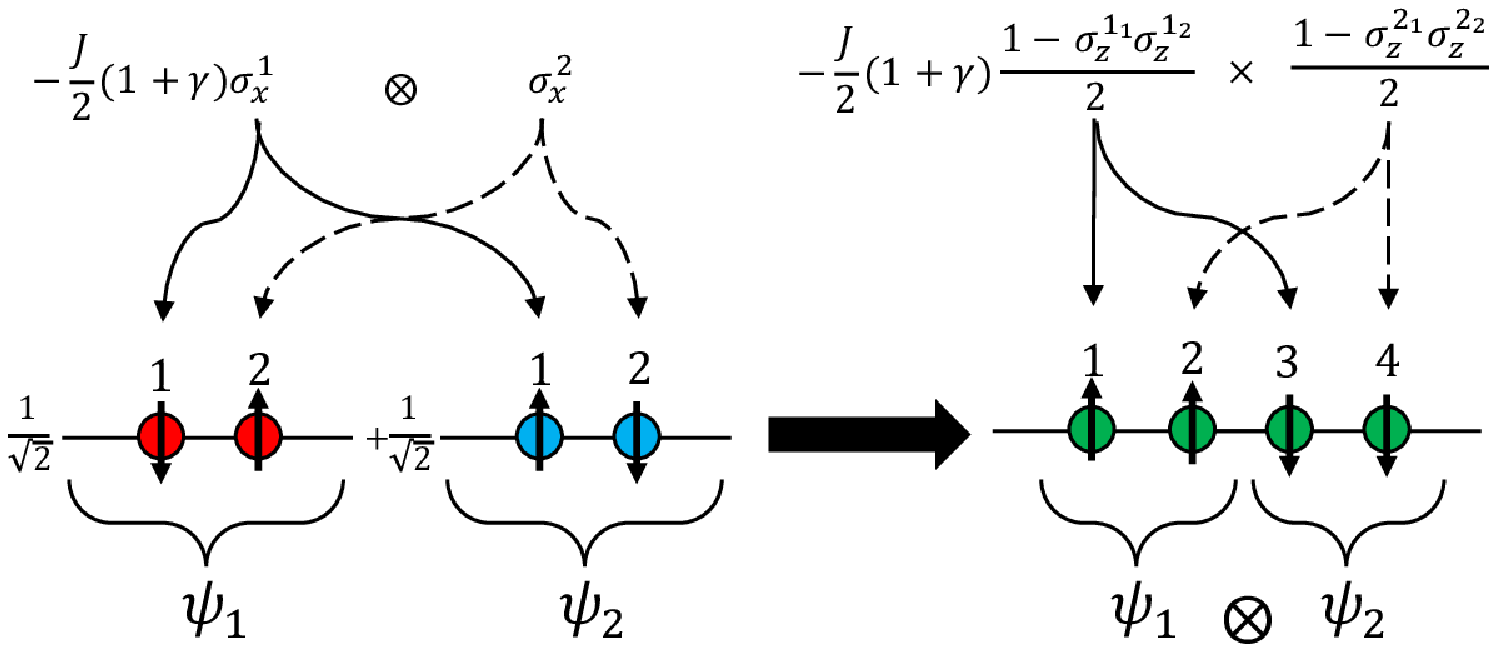}
\caption{The mapped Hamiltonian $-\frac{J}{2}(1+\gamma)\x{1}\x{2}$ between different basis .}
\end{center}
\end{figure}

\begin{equation}
\begin{aligned}
C=&(\frac{1+\z{1}}{2}\frac{1+\z{2}}{2}S'(1)+\frac{1+\z{1}}{2}\frac{1+\z{2}}{2}S'(2))^2\\
&+(\frac{1+\z{1}}{2}\frac{1-\z{2}}{2}S'(1)+\frac{1+\z{1}}{2}\frac{1-\z{2}}{2}S'(2))^2\\
&+(\frac{1-\z{1}}{2}\frac{1+\z{2}}{2}S'(1)+\frac{1-\z{1}}{2}\frac{1+\z{2}}{2}S'(2))^2\\
&+(\frac{1-\z{1}}{2}\frac{1-\z{2}}{2}S'(1)+\frac{1-\z{1}}{2}\frac{1-\z{2}}{2}S'(2))^2 .\\
\end{aligned}
\end{equation}

If $S'(1)=S'(2)=1$, we have:

\begin{equation}
H'=\sum_{j,k}^{j,k \leq 2} H'_{(j,k)}S'(j)S'(k)=H'_{(1,1)}+H'_{(2,2)}+H'_{(1,2)}+H'_{(2,1)} .
\end{equation}

We can write matrix $C$ as:

\[
C=
\begin{bmatrix}
4 & 0 & 0 & 0\\
0 & 2 & 0 & 0\\
0 & 0 & 2 & 0\\
0 & 0 & 0 & 4\\
\end{bmatrix} .
\]

If $-S'(1) =S'(2)=1$, we have:

\begin{equation}
H'=\sum_{j,k}^{j,k \leq 2} H'_{(j,k)}S'(j)S'(k)=H'_{(1,1)}+H'_{(2,2)}-H'_{(1,2)}-H'_{(2,1)}
\end{equation}

We can write $C$ in the matrix format:

\[
C=
\begin{bmatrix}
0 & 0 & 0 & 0\\
0 & 2 & 0 & 0\\
0 & 0 & 2 & 0\\
0 & 0 & 0 & 0\\
\end{bmatrix} .
\]

For $-S'(1)=S'(2)=1$, we here display a procedural representation of our algorithm (we set $B=0.001$, $J=-0.1$ and
$\gamma=0$):

\begin{enumerate}
\item First we choose $\lambda=100$, we get the minimum eigenvalue of $H'-100C$ is $-400$ with $|\Psi\rangle=|0000\rangle$. Thus we get $\sum_mb_m^2=4$ and $\lambda'=0$.

\item We set $\lambda=0$, we get the minimum eigenvalue of $H'+0C$ is $-0.2$ with $|\Psi\rangle=|0110\rangle$. Thus we get $\sum_mb_m^2=2$ and $\lambda'=-0.1$.

\item We set $\lambda=-0.1$, we get the minimum eigenvalue of $H'+0.1C$ is $0$ with $|\Psi\rangle=|0110\rangle$. We stop here and get the minimum eigenvalue of $H$ is $-0.1$
\end{enumerate}

Here we present the result of mapping the above Hamiltonian, Eq.(12), with $B=0.001$, $J=-0.821R^{5/2}e^{-2R}$ and $\gamma=0$\cite{huang2005entanglement}.

\begin{center}
\includegraphics[height=2.2in,]{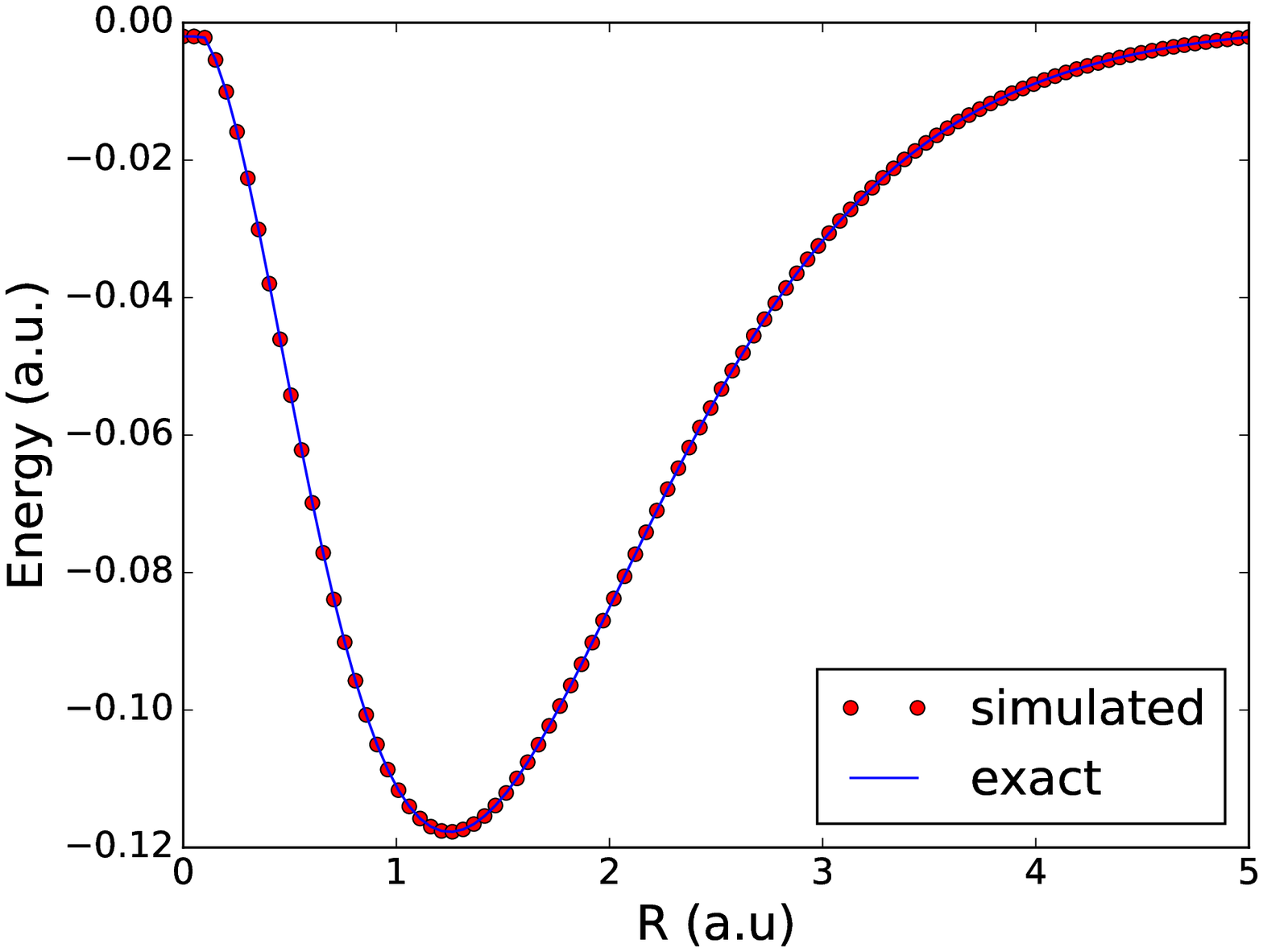}
\captionof{figure}{Comparing the ground state energy from exact (atomic units) of the original Hamiltonian $H$, Eq. (12), as a function of the internuclear distance, $R$, (solid line) with the results of the transformed Hamiltonian $H'$, Eq. (16)(17). }

\end{center}

\subsection*{Reduce Locality of the Transformed Hamiltonian}

Here we present the procedure to reduce the locality of $H'$ from $k$-local to a 2-local Ising-type Hamiltonian.

For $x,\ y,\ z \in \{0,1\}$\cite{bian2013experimental}:

\begin{equation}
xy=z \ \text{iff} \ xy-2xz-2yz+3z=0
\end{equation}

and

\begin{equation}
xy\neq z\ \text{iff} \ xy-2xz-2yz+3z>0
\end{equation}

So the 3-local $x_1x_2x_3$ can be transformed to 2-local by setting $x_4=x_1x_2$:

\begin{equation}
min(x_1x_2x_3)=min(x_4x_3+x_1x_2-2x_1x_4-2x_2x_4+3x_4)\ \ x_1,x_2,x_3,x_4 \in \{0,1\}
\end{equation}

\begin{equation}
min(-x_1x_2x_3)=min(-x_4x_3+x_1x_2-2x_1x_4-2x_2x_4+3x_4)\
 \ x_1,x_2,x_3,x_4 \in \{0,1\}
\end{equation}

We can prove that the $ min(x_1x_2x_3+f(x)=g_1(x))=min(x_4x_3+x_1x_2-2x_1x_4-2x_2x_4+3x_4+f(x)=g_2(x,x_4))$ where $f(x)$ is polynomial of all variables (including $x_1,x_2,x_3$ and other variables, excluding $x_4$).

If there exists $x'$ makes $g_1(x')$ to be minimum, we can always make $g_2(x',x_4)=g_1(x')$ by choosing $x_4=x_1x_2$. Then if there exists $x''$ makes $g_2(x'',x_4)$ to be minimum, then $g_1(x'')\leq g_2(x'')$:

\begin{enumerate}
\item If $x_4=x_1x_2$, $g_1(x'')=x_1x_2x_3+f(x'')=x_4x_3+x_1x_2-2x_1x_4-2x_2x_4+3x_4+f(x'')=g_2(x'',x_4)$.
\item If $x_4\neq x_1x_2$, $g_2(x'',x_4)=x_4x_3+x_1x_2-2x_1x_4-2x_2x_4+3x_4+f(x'')\geq x_4x_3+1+f(x'') \geq x_1x_2x_3+f(x'')=g_1(x'')$ because $x_4x_3-x_1x_2x_3\geq -1$.
\end{enumerate}

Thus, we have $g_1(x')=g_2(x',x_4)\geq g_2(x'',x_4)\geq g_1(x'')\geq g_1(x')$. Thus we have $g_1(x')=g_2(x',x_4)=g_2(x'',x_4)=g_1(x'')$, or all $x$ makes $g_1(x)$ minimum would also makes $g_2(x,x_4)$ minimum and vice versa.

Thus, we obtain:

\begin{equation}
\begin{aligned}
&min(\z{1}\z{2}\prod_{i=3}^n\z{i})\\
&=min(8\frac{1+\z{1}}{2}\frac{1+\z{2}}{2}\frac{1+\prod_{i=3}^n\z{i}}{2}-8\frac{1+\z{1}+\z{2}+\prod_{i=3}^n\z{i}+\z{1}\z{2}+\z{1}\prod_{i=3}^n\z{i}+\z{2}\prod_{i=3}^n\z{i}}{8})\\
&=min(7+\prod_{i=3}^n\z{i}-3\z{1}-3\z{2}+6\z{n+1}+2\prod_{i=3}^{n+1}\z{i}-\z{1}\prod_{i=3}^n\z{i}-\z{2}\prod_{i=3}^n\z{i}-4\z{1}\z{n+1}-4\z{2}\z{n+1}+\z{1}\z{2})
\end{aligned}
\end{equation}

\begin{equation}
\begin{aligned}
&min(-\z{1}\z{2}\prod_{i=3}^n\z{i})\\
&=min(-8\frac{1+\z{1}}{2}\frac{1+\z{2}}{2}\frac{1+\prod_{i=3}^n\z{i}}{2}+8\frac{1+\z{1}+\z{2}+\prod_{i=3}^n\z{i}+\z{1}\z{2}+\z{1}\prod_{i=3}^n\z{i}+\z{2}\prod_{i=3}^n\z{i}}{8})\\
&=min(5-\prod_{i=3}^n\z{i}-\z{1}-\z{2}+2\z{n+1}-2\prod_{i=3}^{n+1}\z{i}+\z{1}\prod_{i=3}^n\z{i}+\z{2}\prod_{i=3}^n\z{i}-4\z{1}\z{n+1}-4\z{2}\z{n+1}+3\z{1}\z{2})
\end{aligned}
\end{equation}

By repeating this, we can reduce the $k$-local in $\sigma_z$ terms to a 2-local Hamiltonian.

\subsection*{Mapping the $H_2$ Hamiltonian to an Ising-type Hamiltonian}

Here, we treat the Hydrogen molecule in a minimal basis STO-6G. By considering the spin functions, the four molecular spin orbitals in $H_2$ are:
\begin{equation}
\ket{\chi_1} = \ket{\Psi_g}\ket{\alpha} = \frac{ \ket{\Psi_{1s}}_1+\ket{\Psi_{1s}}_2}{\sqrt{2(1+S)}}\ket{\alpha}
\end{equation}

\begin{equation}
\ket{\chi_2} = \ket{\Psi_g}\ket{\beta} = \frac{ \ket{\Psi_{1s}}_1+\ket{\Psi_{1s}}_2}{\sqrt{2(1+S)}}\ket{\beta}
\end{equation}

\begin{equation}
\ket{\chi_3} = \ket{\Psi_u}\ket{\alpha} = \frac{ \ket{\Psi_{1s}}_1-\ket{\Psi_{1s}}_2}{\sqrt{2(1-S)}}\ket{\alpha}
\end{equation}

\begin{equation}
\ket{\chi_4} = \ket{\Psi_u}\ket{\beta} = \frac{ \ket{\Psi_{1s}}_1-\ket{\Psi_{1s}}_2}{\sqrt{2(1-S)}}\ket{\beta} ,
\end{equation}

where $\ket{\Psi_{1s}}_1$ and $\ket{\Psi_{1s}}_2$ are the spatial-function for the two atoms respectively, $|\alpha\rangle$, $|\beta\rangle$ are spin up and spin down and $S ={}_{\raisebox{1.535pt}{\scalebox{0.74}{1}}}{\braket{\Psi_{1s}|\Psi_{1s}}}_2$ is the overlap integral\cite{pyquant}. The one and two-electron integrals are giving by

\begin{equation}
h_{ij} = \int{d\vec{r}\chi_i^*(\vec{r})(-\frac{1}{2}\nabla-\frac{Z}{r})\chi_j(\vec{r})}
\end{equation}

\begin{equation}
h_{ijkl} = \int{d\vec{r_1}d\vec{r_2}\chi_{i}^*(\vec{r_1})\chi_{j}^*(\vec{r_2})\frac{1}{r_{12}}\chi_{k}(\vec{r_2})\chi_{l}(\vec{r_1})}
\end{equation}

Thus, we can write the second-quantization Hamiltonian of $H_2$:
\begin{equation}
\begin{aligned}
H_{H_2} &= h_{00}a_0^\dagger a_0 + h_{11}a_1^\dagger a_1+ h_{22}a_2^\dagger a_2 + h_{33}a_3^\dagger a_3+
h_{0110}a_0^\dagger a_1^\dagger a_1 a_0 +h_{2332}a_2^\dagger a_3^\dagger a_3 a_2 + h_{0330}a_0^\dagger a_3^\dagger a_3 a_0\\
&+ h_{1221} a_1^\dagger a_2^\dagger a_2 a_1 + (h_{0220} - h_{0202})a_0^\dagger a_2^\dagger a_2 a_0
+ (h_{1331}-h_{1313})a_1^\dagger a_3^\dagger a_3 a_1 \\
&+h_{0132}(a_0^\dagger a_1^\dagger a_3 a_2 + a_2^\dagger a_3^\dagger a_1 a_0)+
h_{0312}(a_0^\dagger a_3^\dagger a_1 a_2 + a_2^\dagger a_1^\dagger a_3 a_0)
\end{aligned}
\end{equation}

By using Bravyi-Kitaev transformation\cite{seeley2012bravyi}, we have:
\begin{equation}
\label{eqn3_2}
\begin{aligned}
&a_0^{\dagger}=\frac{1}{2}\sigma_x^3\sigma_x^1(\sigma_x^0- i\sigma_y^0)\quad
a_0=\frac{1}{2}\sigma_x^3\sigma_x^1(\sigma_x^0+i\sigma_y^0)\quad
a_1^{\dagger}=\frac{1}{2}(\sigma_x^3\sigma_x^1\sigma_z^0- i\sigma_x^3\sigma_y^1)\quad
a_1=\frac{1}{2}(\sigma_x^3\sigma_x^1\sigma_z^0+i\sigma_x^3\sigma_y^1)\\
&a_2^{\dagger}=\frac{1}{2}\sigma_x^3(\sigma_x^2- i\sigma_y^2)\sigma_z^1\quad
a_2=\frac{1}{2}\sigma_x^3(\sigma_x^2+ i\sigma_y^2)\sigma_z^1\quad
a_3^{\dagger}=\frac{1}{2}(\sigma_x^3\sigma_z^2\sigma_z^1- i\sigma_y^3)\quad
a_3=\frac{1}{2}(\sigma_x^3\sigma_z^2\sigma_z^1+ i\sigma_y^3) .
\end{aligned}
\end{equation}

Thus, the Hamiltonian of $H_2$ takes the following form:

\begin{equation}
\label{eqn3_2}
\begin{aligned}
H_{H_2}&=f_0{\bf 1}+f_1\sigma_z^0+f_2\sigma_z^1+f_3\sigma_z^2+f_1\sigma_z^0\sigma_z^1+f_4\sigma_z^0\sigma_z^2+f_5\sigma_z^1\sigma_z^3+f_6\sigma_x^0\sigma_z^1\sigma_x^2+f_6\sigma_y^0\sigma_z^1\sigma_y^2\\
&+f_7\sigma_z^0\sigma_z^1\sigma_z^2+f_4\sigma_z^0\sigma_z^2\sigma_z^3+f_3\sigma_z^1\sigma_z^2\sigma_z^3+f_6\sigma_x^0\sigma_z^1\sigma_x^2\sigma_z^3+f_6\sigma_y^0\sigma_z^1\sigma_y^2\sigma_z^3+f_7\sigma_z^0\sigma_z^1\sigma_z^2\sigma_z^3 .
\end{aligned}
\end{equation}

We can utilize the symmetry that qubits 1 and 3 never flip to reduce the Hamiltonian to the following form which just acts on only two qubits:

\begin{equation}
H_{H_2}=g_0{\bf 1}+g_1\sigma_z^0+g_2\sigma_z^1+g_3\sigma_z^0\sigma_z^1+g_4\sigma_x^0\sigma_x^1+g_4\sigma_y^0\sigma_y^1=g_0{\bf 1}+H_0
\end{equation}

\begin{equation}
g_0=f_0\ g_1=2f_1\ g_2=2f_3\ g_3=2(f_4+f_7)\ g_4=2f_6
\end{equation}

\begin{equation}
\begin{aligned}
&g_0=1.0h_{00} + 0.5h_{0000} - 0.5h_{0022}+ 1.0h_{0220} + 1.0h_{22} + 0.5h_{2222} + 1.0/R\\
&g_1=-1.0h_{00} - 0.5h_{0000} + 0.5h_{0022} - 1.0h_{0220}\\
&g_2=0.5h_{0022} - 1.0h_{0220} - 1.0h_{22} - 0.5h_{2222}\\
&g_3=-1.0h_{00} - 0.5h_{0000} + 0.5h_{0022} - 1.0h_{0220}\\
&g_4=0.5h_{0022}
\end{aligned}
\end{equation}

Where $\left\{g_i\right\}$ depends on the fixed bond length of the molecule. In Table I, we present the numerical values of $\left\{g_i\right\}$ as a function of the internuclear distance in the minimal basis set STO-6G.

By applying the mapping method described above, we can get the Hamiltonian $H'$ consisting of only $\sigma_z$ (where $i_1$ and $i_2$ means the 1 and 2 qubits of $i^{th}$ 2 qubits):

\begin{equation}
\begin{aligned}
H'&=\sum_{i}g_1\z{i_1}+g_2\z{i_2}+g_3\z{i_1}\z{i_2}+\sum_{i\neq j}[g_1\frac{(\z{i_1}+\z{j_1})(1+\z{i_2}\z{j_2})}{4}S'(i)S'(j)\\
&+g_2\frac{(\z{i_2}+\z{(j,2}))(1+\z{i_1}\z{j_1})}{4}S'(i)S'(j)+g_3\frac{(\z{i_1}+\z{j_1})(\z{i_2}+\z{j_2})}{4}S'(i)S'(j)\\
&+g_4\frac{(1+\z{i_1}\z{i_1})(1+\z{i_2}\z{j_2})}{4}S'(i)S'(j)+g_4\frac{(\z{i_1}-\z{i_1})(\z{j_2}-\z{i_2})}{4}S'(i)S'(j)]
\end{aligned}
\end{equation}

According to the scheme for reducing locality, if we want to reduce $H'-\lambda C$, we can reduce $H'$ and $C$ separately. By applying the method for reducing locality, we can get a 2-local Ising-type Hamiltonian, $H''$. Here we show the example of all signs are positive, where $a_{ij}^k$, $k=1,2,3,...,6$ are the index of the new qubits we introduce to reduce locality. 

\begin{equation}
\begin{aligned}
H''&=\sum_{i}g_1\z{i_1}+g_2\z{i_2}+g_3\z{i_1}\z{i_2}+\sum_{i \neq j}[g_1\frac{\z{i_1}+\z{j_1}}{4}\\
&+g_2\frac{\z{i_2}+\z{j_2}}{4}+g_3\frac{(\z{i_1}+\z{j_1})(\z{i_2}+\z{j_2})}{4}\\
&+g_4\frac{1+\z{i_1}\z{j_1}+\z{i_2}\z{j_2}}{4}+g_4\frac{(\z{i_1}-\z{j,1)})(\z{i_2}-\z{j_2})}{4}\\
&+g_4\frac{7+\z{i_2}\z{j_2}-3\z{i_1}-3\z{j_1}+6\z{a_{ij}^1}-4\z{i_1}\z{a_{ij}^1}-4\z{j_1}\z{a_{ij}^1}+\z{i_1}\z{j_1}}{4}\\
&+(g_1-g_4)\frac{14+\z{i_1}+\z{j_1}-6(\z{i_2}+\z{j_2})+6(\z{a_{ij}^2}+\z{a_{ij}^3})+2(\z{i_1}\z{a_{ij}^2}+\z{j_1}\z{a_{ij}^3})}{4}\\
&+(g_1-g_4)\frac{-4(\z{i_2}+\z{j_2})(\z{a_{ij}^2}+\z{a_{ij}^3})-(\z{i_1}+\z{j_1})(\z{i_2}+\z{j_2})+2\z{i_2}\z{j_2}}{4}\\
&+g_2\frac{10-\z{i_2}-\z{j_2}-2(\z{i_1}+\z{j_1})+2(\z{a_{ij}^4}+\z{a_{ij}^5})-2(\z{i_2}\z{a_{ij}^4}+\z{j_2}\z{a_{ij}^5})}{4}\\
&+g_2\frac{-4(\z{i_1}+\z{j_1})(\z{a_{ij}^4}+\z{a_{ij}^5})+(\z{i_1}+\z{j_1})(\z{i_2}+\z{j_2})+6\z{i_1}\z{j_1}}{4}\\
&+g_4\frac{7+\z{a_{ij}^1}-3\z{i_2}-3\z{j_2}+6\z{a_{ij}^6}+2\z{a_{ij}^1}\z{a_{ij}^6}}{2}\\
&+g_4\frac{-\z{i_2}\z{a_{ij}^1}-\z{j_2}\z{a_{ij}^1}-4\z{i_2}\z{a_{ij}^6}-4\z{j_2}\z{a_{ij}^6}+\z{i_2}\z{j_2}}{2}]
\end{aligned}
\end{equation}

We can write the corresponding count term $C$ as:
\begin{equation}
C=\sum_{\pm}(\sum_i\frac{(1\pm\z{i_1})(1\pm \z{i_2})}{4})^2
\end{equation}

By applying the method of reducing locality, 2-local corresponding count term is:

\begin{equation}
\begin{aligned}
C'&=\frac{1}{4}\{\sum_{i \neq j}[7+\z{i_2}\z{j_2}-3\z{i_1}-3\z{j_1}+6\z{a_{ij}^7}-4\z{i_1}\z{a_{ij}^7}-4\z{j_1}\z{a_{ij}^7}+\z{i_1}\z{j_1}\\
&+2(7+\z{a_{ij}^7}-3\z{i_2}-3\z{j_2}+6\z{a_{ij}^8}+2\z{a_{ij}^1}\z{a_{ij}^8})\\
&+2(-\z{i_2}\z{a_{ij}^7}-\z{j_2}\z{a_{ij}^7}-4\z{i_2}\z{a_{ij}^8}-4\z{j_2}\z{a_{ij}^8}+\z{i_2}\z{j_2})\\
&+14+\z{i_1}+\z{j_1}-6(\z{i_2}+\z{j_2})+6(\z{a_{ij}^9}+\z{a_{ij}^10})+2(\z{i_1}\z{a_{ij}^9}+\z{j_1}\z{a_{ij}^10})\\
&-4(\z{i_2}+\z{j_2})(\z{a_{ij}^9}+\z{a_{ij}^10})-(\z{i_1}+\z{j_1})(\z{i_2}+\z{j_2})+2\z{i_2}\z{j_2}\\
&+1+\z{i_1}\z{j_1}+\z{i_2}\z{j_2}]\}+\sum_{i}\frac{(1+\z{i_1})+(1+\z{i_2})}{4}
\end{aligned}
\end{equation}

As stated in \S {\bf Reduce Locality}, a change in the locality would not change the state when calculating the ground state energy. Thus we can still use $C$ on certain qubits to calculate $\sum_ia_i^2$, and the algorithm we present above can still be used for the reduced Hamiltonian.

In Figure 1, we show our results from the transformed Ising-type Hamiltonian of $2\times2$ to $4\times2$ qubits compared with the exact numerical values. By increasing the number of qubits via $r=2$, we increased the accuracy and the result matches very well with exact results.

\begin{figure}[H]
\begin{minipage}[t]{0.5\linewidth}
\centering
\includegraphics[width=3.5in]{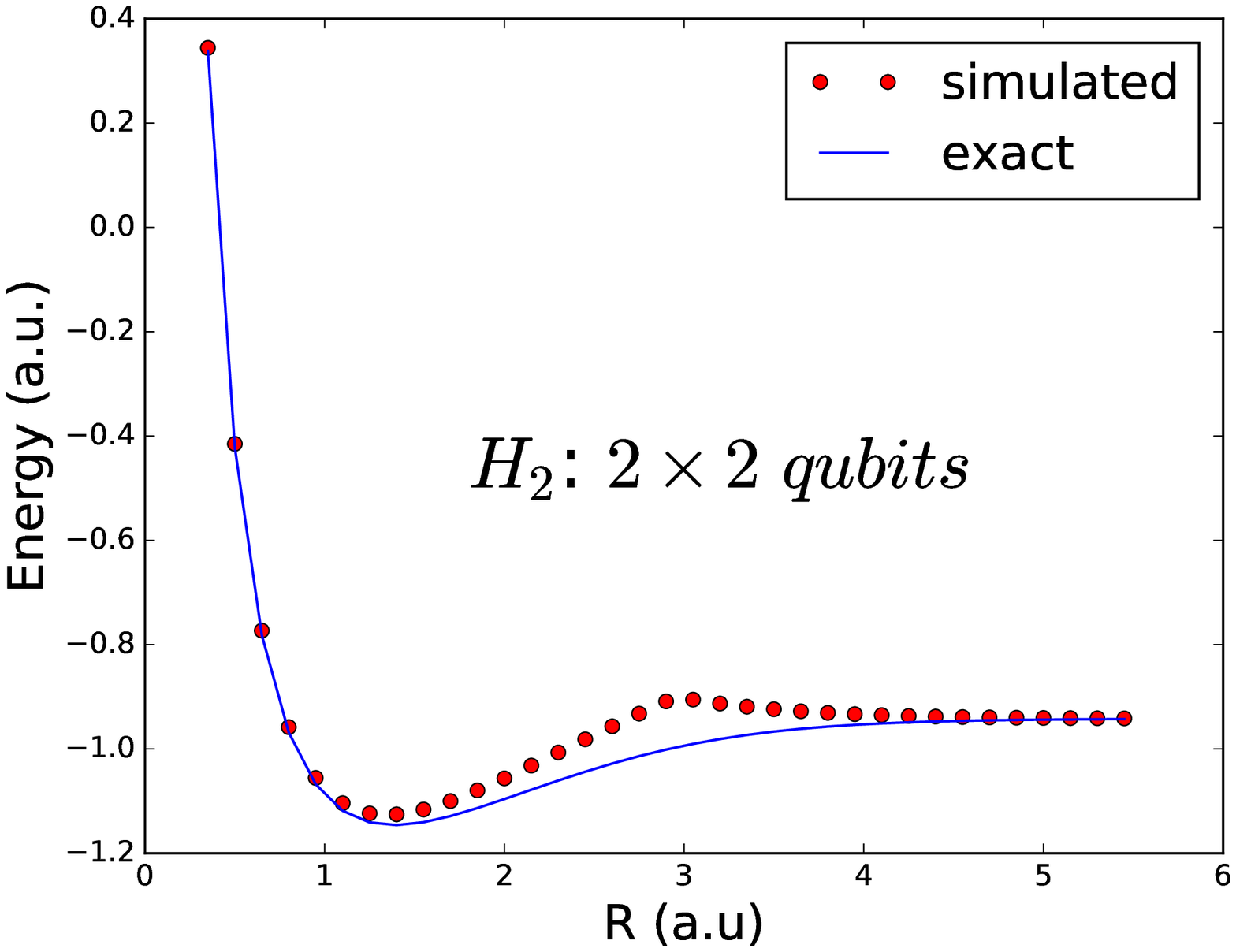}
\label{fig:side:a}
\end{minipage}%
\begin{minipage}[t]{0.5\linewidth}
\centering
\includegraphics[width=3.5in]{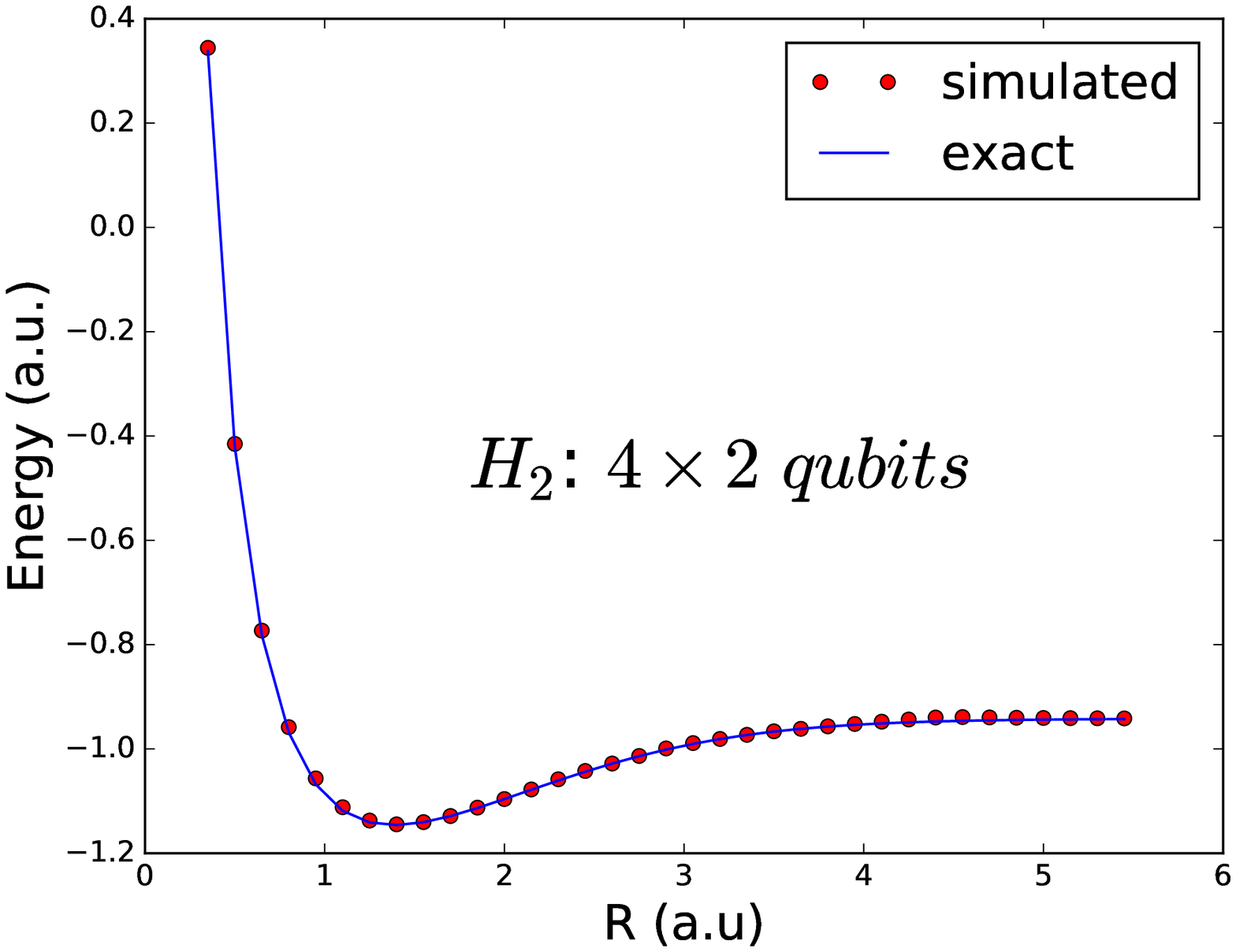}
\label{fig:side:b}
\end{minipage}

\caption{Results of  the simulated transformed Ising-type Hamiltonian with $2\times 2$ qubits and $4 \times 2$ qubits compared with the exact numerical results for ground state of H$_2$ molecule.}
\end{figure}

\subsection*{Mapping the Hamiltonian for He$_2$ Molecule to the Ising-type Hamiltonian}

As shown above for transforming the Hamiltonian associated with the H$_2$ molecule, we repeat the procedure for the Helium molecule in a minimal basis STO-6G using Jordanâ-Wigner transformation.

The molecular spin Hamiltonian has the form: 

\begin{equation}
\begin{aligned}
&H_{He_2}=f_0{\bf 1}+f_1\z{1}+f_2\z{2}+f_3\z{3}+f_4\z{4}+f_5\z{1}\z{2}+f_6\z{1}\z{3}+f_7\z{1}\z{4}+f_8\z{2}\z{3}+f_8\z{2}\z{4}\\
&+f_9\z{3}\z{4}+f_{10}\x{1}\x{2}\y{3}\y{4}+f_{11}\x{1}\y{2}\y{3}\x{4}+f_{12}\y{1}\x{2}\x{3}\y{4}+f_{13}\y{1}\y{2}\x{3}\x{4} .
\end{aligned}
\end{equation}

The set of parameters ${f_i}$ are related to the one and two electron integrals:

\begin{equation}
\begin{aligned}
&f_0=1.0h_{00} + 0.25h_{0000} - 0.5h_{0022} + 1.0h_{0220} + 1.0h_{22} + 0.25h_{2222} + 4.0/R\\
&f_1=-0.5h_{00} - 0.25h_{0000} + 0.25h_{0022} - 0.5h_{0220}\\
&f_2=-0.25h_{0000} + 0.25h_{0022} - 0.5h_{0220} - 0.5h_{00}\\
&f_3=0.25h_{0022} - 0.5h_{0220} - 0.5h_{22} - 0.25h_{2222}\\
&f_4=0.25h_{0022} - 0.5h_{0220} - 0.25h_{2222} - 0.5h_{22}\\
&f_5=0.25h_{0000}\quad f_6=-0.25h_{0022} + 0.25h_{0220}\quad f_7=0.25h_{0220}\\
&f_8=0.25h_{0220}\quad f_9=0.25h_{2222}\quad f_{10}=-0.25h_{0022}\\ &f_{11}=0.25h_{0022}\quad f_{12}=0.25h_{0022}\quad f_{13}=-0.25h_{0022} . \\
\end{aligned}
\end{equation}

We can also use the mapping and reduction of locality as before to get the final Ising Hamiltonian. Here we just present the mapping result of some terms for illustration.

For $\z{1}$, the Hamiltonian between different basis can be mapped as:
\begin{equation}
\begin{aligned}
\sum_{i \neq j}\frac{(\z{i_1}+\z{j_1})(1+\z{i_2}\z{j_2})(1+\z{i_3}\z{j_3})(1+\z{j_4}\z{i_4})}{16}S'(i)S'(j) .
\end{aligned}
\end{equation}

For $\z{1}\z{2}$, the Hamiltonian between different basis can be mapped as:
\begin{equation}
\begin{aligned}
\sum_{i \neq j}\frac{(\z{i_1}+\z{j_1})(\z{i_2}+\z{j_2})(1+\z{i_3}\z{j_3})(1+\z{j_4}\z{i_4})}{16}S'(i)S'(j) .
\end{aligned}
\end{equation}

For $\x{1}\x{2}\y{3}\y{4}$, the Hamiltonian between different basis can be mapped as:

\begin{equation}
\begin{aligned}
\sum_{i \neq j}\frac{(1-\z{i_1}\z{j_1})(1-\z{i_2}\z{j_2})(\z{i_3}-\z{j_3})(\z{j_4}-\z{i_4})}{16}S'(i)S'(j) .
\end{aligned}
\end{equation}

By reducing locality, we can get 2-local Ising-type Hamiltonian. However, even just for $\x{1}\x{2}\y{3}\y{4}$, the final 2-local Ising Hamiltonian would have about 1000 terms. 

\begin{center}
\includegraphics[height=3in,]{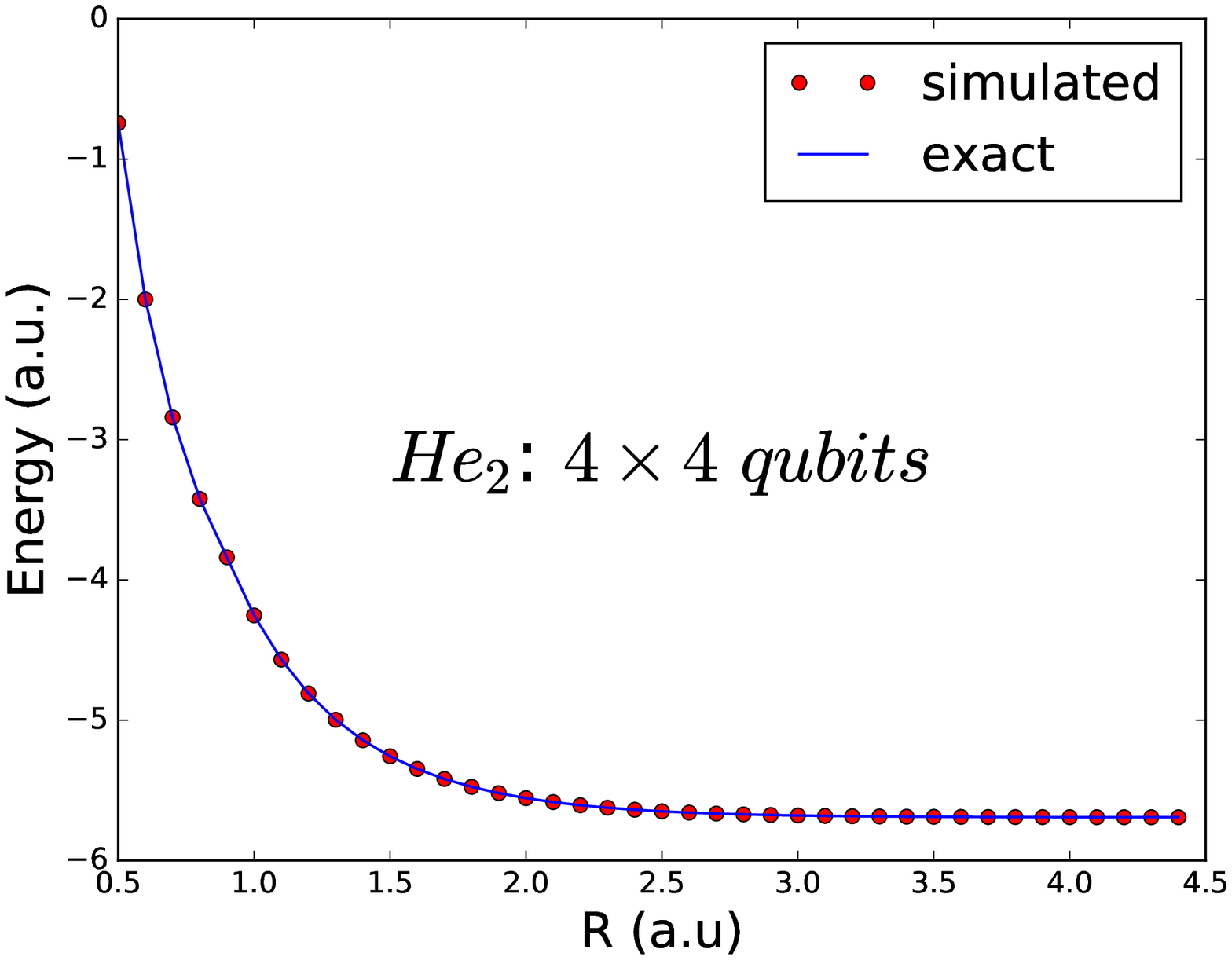}
\captionof{figure}{Comparing the exact results for the ground state electronic energy of He$_2$ molecule as a function of internuclear distance, $R$, (solid line) with the simulated transformed Hamiltonian (dots) for $8$ qubits under the minimal basis set STO-6G}
\end{center}

\subsection*{Mapping the Hamiltonian of HeH$^+$ to an Ising Hamiltonian}

Similar to H$_2$ and He$_2$ molecules, next we treat HeH$^+$ molecule in the minimal basis STO-6G using Jordanâ-Wigner transformation. Using the technique defined above \cite{moll2016optimizing} we can reduce the locality to:

\begin{equation}
\begin{aligned}
&H_{HeH^+}=f_0{\bf 1}+f_1\z{1}+f_2\z{2}+f_3\x{1}+f_4\z{1}\z{2}+f_5\x{1}\z{2}+f_6\z{1}\x{2}+f_7\x{1}\x{2}+f_8\y{1}\y{2} .
\end{aligned}
\end{equation}

The set of parameters ${f_i}$ are related to the one and two electron integrals:

\begin{equation}
\begin{aligned}
&f_0=1.0h_{00} + 0.25h_{0000} + 0.5h_{0220} + 1.0h_{22} + 0.25h_{2222} + 2.0/R\\
&f_1=-0.25h_{0000 }+ 0.5h_{0220 }- 0.25h_{2222}\\
&f_2=0.25h_{0000 }+ 0.5h_{00 }- 0.25h_{2222 }- 0.5h_{22}\\
&f_3=-0.25h_{0002 }- 0.25h_{0020 }+ 0.5h_{0222}\\
&f_4=-0.5h_{00 }- 0.25h_{0000 }+ 0.5h_{22 }+ 0.25h_{2222}\\
&f_5=-0.25h_{0002 }- 0.25h_{0020 }- 0.5h_{0222}\\
&f_6=-1.0h_{0022}\\
&f_7=0.25h_{0002 }+ 0.25h_{0020 }- 0.5h_{0222}\\
&f_8=0.25h_{0002 }+ 0.25h_{0020 }+ 0.5h_{0222} .
\end{aligned}
\end{equation}

We can also use the mapping and reducing locality as before to get the final Ising Hamiltonian. Again, here we just present the mapping result of some terms for illustration:

For $\z{1}\z{2}$, the Hamiltonian between different basis can be mapped as:
\begin{equation}
\begin{aligned}
\sum_{i \neq j}\frac{(\z{i_1}+\z{j_1})(\z{i_2}+\z{j_2})}{4}S'(i)S'(j) .
\end{aligned}
\end{equation}

For $\z{1}\x{2}$, the Hamiltonian between different basis can be mapped as:
\begin{equation}
\begin{aligned}
\sum_{i \neq j}\frac{(\z{i_1}+\z{j_1})(1-\z{i_2}\z{j_2})}{4}S'(i)S'(j) .
\end{aligned}
\end{equation}

And if the coefficient of mapping term is positive, we can get the 2-local term as:

\begin{equation}
\begin{aligned}
&\sum_{i\neq j}(\frac{\z{i_1}+\z{j_1}}{4}+\frac{14+\z{i_1}+\z{j_1}-6(\z{i_2}+\z{j_2})+6(\z{a_{ij}^1}+\z{a_{ij}^2})+2(\z{i_1}\z{a_{ij}^1}+\z{j_1}\z{a_{ij}^2})}{4}\\
&+\frac{-4(\z{i_2}+\z{j_2})(\z{a_{ij}^1}+\z{a_{ij}^2})-(\z{i_1}+\z{j_1})(\z{i_2}+\z{j_2})+2\z{i_2}\z{j_2}}{4}) .\\
\end{aligned}
\end{equation}

\begin{figure}[H]
\begin{center}
\includegraphics[width=3.5in]{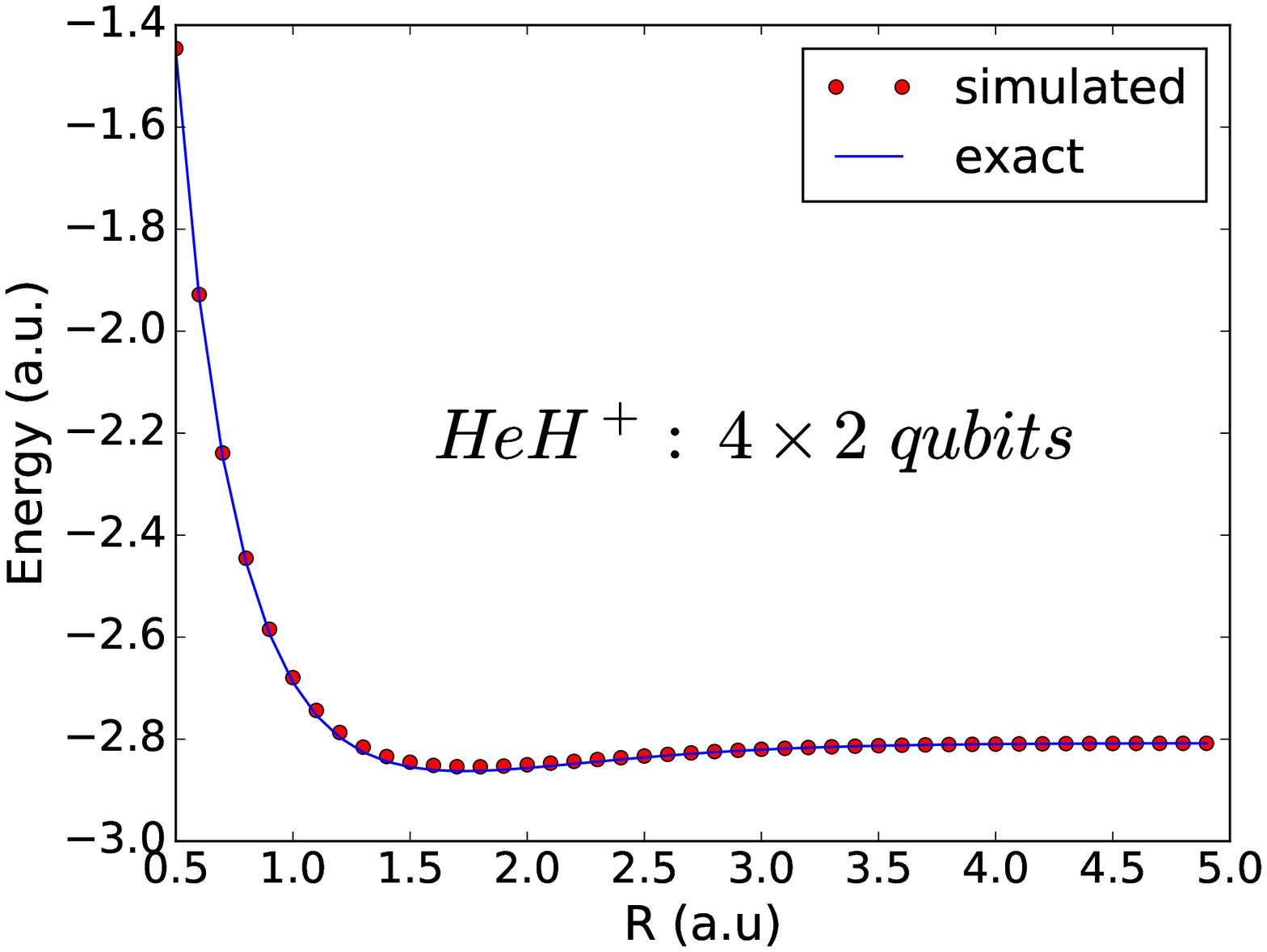}
\caption{A comparison of the exact results for the ground state electronic energy of HeH$^+$ molecule as a function of internuclear distance $R$ (solid line) with the simulated transformed Hamiltonian (dots) for $8$ qubits in the minimal basis set STO-6G.}
\end{center}
\end{figure}

\subsection*{Mapping the molecular Hamiltonian of LiH to an Ising Hamiltonian}

Similar to H$_2$ and other molecules, next we treat $LiH$ molecule with 4-electrons in a minimal basis STO-6G and use of Jordanâ-Wigner transformation. Using the technique defined above \cite{bravyi2017tapering} we can reduce the locality to a Hamiltonian with $558$ terms on $8$ qubits. We just use $16$ qubits for the simulations. 

As in simulating of $H_2$ and $HeH^+$, if we simulate $LiH$ with more qubits we should get more accurate result. Because of computer resources, we run the simulations as shown in Fig (7) with only 16-qubits.

\begin{figure}[H]
\begin{center}
\includegraphics[width=4in]{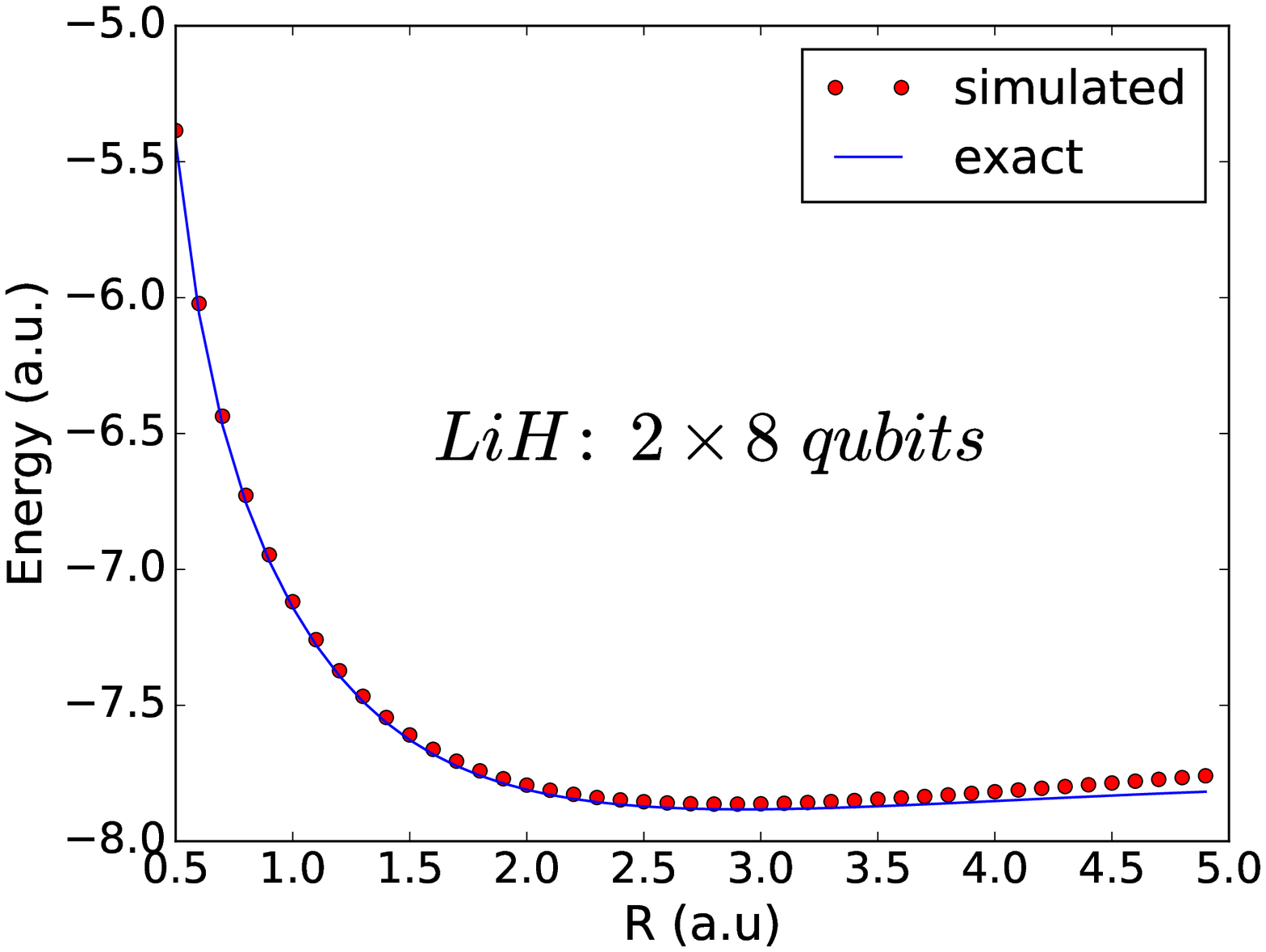}
\caption{Results for simulating with $8 \times 2$ qubits.}
\end{center}
\end{figure}

\begin {table}
\caption {Comparing the Exact Ground Energy (a.u.) and the Simulated Ground Energy (a.u.) (simulated by $4\times 2$ qubits) as a function of the inter-molecule distance R (a.u.)} \label{tab:title} 
\begin{center}
\begin{tabular}{l*{7}{c}} R & $g_0$ & $g_1$ & $g_2$ & $g_3$ & $g_4$ & Exact & Simulated \\ \hline 0.6 & 1.5943 & 0.5132 & -1.1008 & 0.6598 & 0.0809 & -0.5617 & -0.5703 \\0.65 & 1.4193 & 0.5009 & -1.0366 & 0.6548 & 0.0813 & -0.6785 & -0.6877 \\0.7 & 1.2668 & 0.4887 & -0.9767 & 0.6496 & 0.0818 & -0.7720 & -0.7817 \\0.75 & 1.1329 & 0.4767 & -0.9208 & 0.6444 & 0.0824 & -0.8472 & -0.8575 \\0.8 & 1.0144 & 0.465 & -0.8685 & 0.639 & 0.0829 & -0.9078 & -0.9188 \\0.85 & 0.909 & 0.4535 & -0.8197 & 0.6336 & 0.0835 & -0.9569 & -0.9685 \\0.9 & 0.8146 & 0.4422 & -0.774 & 0.6282 & 0.084 & -0.9974 & -1.0088 \\0.95 & 0.7297 & 0.4313 & -0.7312 & 0.6227 & 0.0846 & -1.0317 & -1.0415 \\1.0 & 0.6531 & 0.4207 & -0.691 & 0.6172 & 0.0852 & -1.0595 & -1.0678 \\1.05 & 0.5836 & 0.4103 & -0.6533 & 0.6117 & 0.0859 & -1.0820 & -1.0889 \\1.1 & 0.5204 & 0.4003 & -0.6178 & 0.6061 & 0.0865 & -1.0999 & -1.1056 \\1.15 & 0.4626 & 0.3906 & -0.5843 & 0.6006 & 0.0872 & -1.1140 & -1.1186 \\1.2 & 0.4098 & 0.3811 & -0.5528 & 0.5951 & 0.0879 & -1.1249 & -1.1285 \\1.25 & 0.3613 & 0.37200 & -0.523 & 0.5897 & 0.0886 & -1.1330 & -1.1358 \\1.3 & 0.3167 & 0.3631 & -0.4949 & 0.5842 & 0.0893 & -1.1389 & -1.1409 \\1.35 & 0.2755 & 0.3546 & -0.4683 & 0.5788 & 0.09 & -1.1427 & -1.1441 \\1.4 & 0.2376 & 0.3463 & -0.4431 & 0.5734 & 0.0907 & -1.1448 & -1.1457 \\1.45 & 0.2024 & 0.3383 & -0.4192 & 0.5681 & 0.0915 & -1.1454 & -1.1459 \\1.5 & 0.1699 & 0.3305 & -0.3966 & 0.5628 & 0.0922 & -1.1448 & -1.145 \\1.55 & 0.1397 & 0.32299 & -0.3751 & 0.5575 & 0.09300 & -1.1431 & -1.1432 \\1.6 & 0.1116 & 0.3157 & -0.3548 & 0.5524 & 0.0938 & -1.1404 & -1.1405 \\1.65 & 0.0855 & 0.3087 & -0.3354 & 0.5472 & 0.0946 & -1.1370 & -1.1371 \\1.7 & 0.0612 & 0.3018 & -0.317 & 0.5422 & 0.0954 & -1.1329 & -1.1332 \\1.75 & 0.0385 & 0.2952 & -0.2995 & 0.5371 & 0.0962 & -1.1281 & -1.1287 \\1.8 & 0.0173 & 0.2888 & -0.2829 & 0.5322 & 0.09699 & -1.1230 & -1.1239 \\1.85 & -0.0023 & 0.2826 & -0.267 & 0.5273 & 0.0978 & -1.1183 & -1.1187 \\1.9 & -0.0208 & 0.2766 & -0.252 & 0.5225 & 0.0987 & -1.1131 & -1.1133 \\1.95 & -0.0381 & 0.2707 & -0.2376 & 0.5177 & 0.0995 & -1.1076 & -1.1077 \\2.0 & -0.0543 & 0.2651 & -0.2238 & 0.513 & 0.1004 & -1.1018 & -1.1019 \\2.05 & -0.0694 & 0.2596 & -0.2108 & 0.5084 & 0.1012 & -1.0958 & -1.0961 \\2.1 & -0.0837 & 0.2542 & -0.1983 & 0.5039 & 0.1021 & -1.0895 & -1.0901 \\2.15 & -0.0969 & 0.249 & -0.1863 & 0.4994 & 0.10300 & -1.0831 & -1.0842 \\2.2 & -0.1095 & 0.244 & -0.1749 & 0.495 & 0.1038 & -1.0765 & -1.0782 \\2.25 & -0.1213 & 0.2391 & -0.1639 & 0.4906 & 0.1047 & -1.0699 & -1.0723 \\2.3 & -0.1323 & 0.2343 & -0.1536 & 0.4864 & 0.1056 & -1.0630 & -1.0664 \\2.35 & -0.1427 & 0.2297 & -0.1436 & 0.4822 & 0.1064 & -1.0581 & -1.0605 \\2.4 & -0.1524 & 0.2252 & -0.1341 & 0.478 & 0.1073 & -1.0533 & -1.0548 \\2.45 & -0.1616 & 0.2208 & -0.125 & 0.474 & 0.1082 & -1.0484 & -1.0492 \\2.5 & -0.1703 & 0.2165 & -0.1162 & 0.47 & 0.109 & -1.0433 & -1.0437 \\2.55 & -0.1784 & 0.2124
 & -0.1079 & 0.466 & 0.1099 & -1.0382 & -1.0383 \\2.6 & -0.1861 & 0.2083 & -0.0999 & 0.4622 & 0.1108 & -1.0330 & -1.0331 \\2.65 & -0.1933 & 0.2044 & -0.0922 & 0.4584 & 0.1117 & -1.0278 & -1.028 \\2.7 & -0.2001 & 0.2006 & -0.0848 & 0.4547 & 0.1125 & -1.0227 & -1.0231 \\2.75 & -0.2064 & 0.1968 & -0.0778 & 0.451 & 0.1134 & -1.0175 & -1.0184 \\2.8 & -0.2125 & 0.1932 & -0.071 & 0.4475 & 0.1142 & -1.0124 & -1.0139 \\2.85 & -0.2182 & 0.1897 & -0.0646 & 0.4439 & 0.1151 & -1.0072 & -1.0095 \\2.9 & -0.2235 & 0.1862 & -0.0584 & 0.4405 & 0.1159 & -1.0021 & -1.0053 \\2.95 & -0.2286 & 0.1829 & -0.0524 & 0.4371 & 0.1168 & -0.9988 & -1.0013 \\3.0 & -0.2333 & 0.1796 & -0.0467 & 0.4338 & 0.1176 & -0.9958 & -0.9974 \\3.05 & -0.2378 & 0.1764 & -0.0413 & 0.4305 & 0.1184 & -0.9928 & -0.9938 \\3.1 & -0.2421 & 0.1733 & -0.0360 & 0.4273 & 0.1193 & -0.9898 & -0.9903 \\\end{tabular}
\end{center}
\end{table}
\end{document}